%% file: main-cycle-sampling.tex
\documentclass[%
 reprint,
nofootinbib,
 amsmath,amssymb,
 aps,
 prx,
 longbibliography,
 showkeys,
floatfix,
]{revtex4-2}

\usepackage{subcaption}
\usepackage{lua-visual-debug}

\usepackage{algpseudocode}
\usepackage{float}
\makeatletter
\let\float@end\float@end@ltx
\let\newfloat\newfloat@ltx
\makeatother
\usepackage{algorithm}
\makeatletter
\def\fnum@algorithm{\fname@algorithm~\thealgorithm}
\makeatother
\usepackage[utf8]{inputenc}
\usepackage[T1]{fontenc}
\usepackage{hyperref} 
\usepackage{url}
\usepackage{booktabs}
\usepackage{amsfonts}
\usepackage{nicefrac}
\usepackage{microtype}
\usepackage{xcolor}
\usepackage{graphicx}
\usepackage{amsthm}
\usepackage[nameinlink,capitalise,noabbrev]{cleveref}
\usepackage{caption,placeins}
\newcommand{\panellabel}[1]{\par\smallskip{\centering\textbf{\textsf{#1}}\par}}
\usepackage{import}
\usepackage{tikz}
\usepackage{mathtools}
\usepackage{bbm}
\usetikzlibrary{positioning}

\hypersetup{
  hidelinks,
  pdftitle={Efficiently Estimating Cycle Statistics on Graphs via Spanning Trees},
  pdfauthor={Josef Hoppe and Michael T. Schaub},
  pdfkeywords={Simple cycles, uniform spanning trees, inclusion-probability estimation, cycle counting, signed networks}
}

\newtheorem{theorem}{Theorem}
\newtheorem{remark}{Remark}

\newcommand{\erdosrenyi}{Erd\H{o}s--R\'enyi}

\DeclareMathOperator{\neigh}{Neigh}
\DeclareMathOperator{\lca}{lca}
\DeclareMathOperator{\depth}{depth}

\newcommand{\repolink}{\href{https://github.com/josefhoppe/cycle-statistics/}{\texttt{github.com/\allowbreak{}josefhoppe/cycle-\allowbreak{}statistics}}}

\date{\today}

\begin{document}

\title{Efficiently Estimating Cycle Statistics on Graphs via Spanning Trees}
\author{Josef Hoppe}
\author{Michael T. Schaub}%
 \email{\{hoppe, schaub\}@netsci.rwth-aachen.de}
\affiliation{%
 RWTH Aachen University
}%

\begin{abstract}
    \input{00-abstract.tex}
\end{abstract}


\maketitle

\input{01-introduction.tex}

\input{02-background}

\input{03-algorithm}

\begin{figure*}
  \centering
  \includegraphics[width=.8\linewidth]{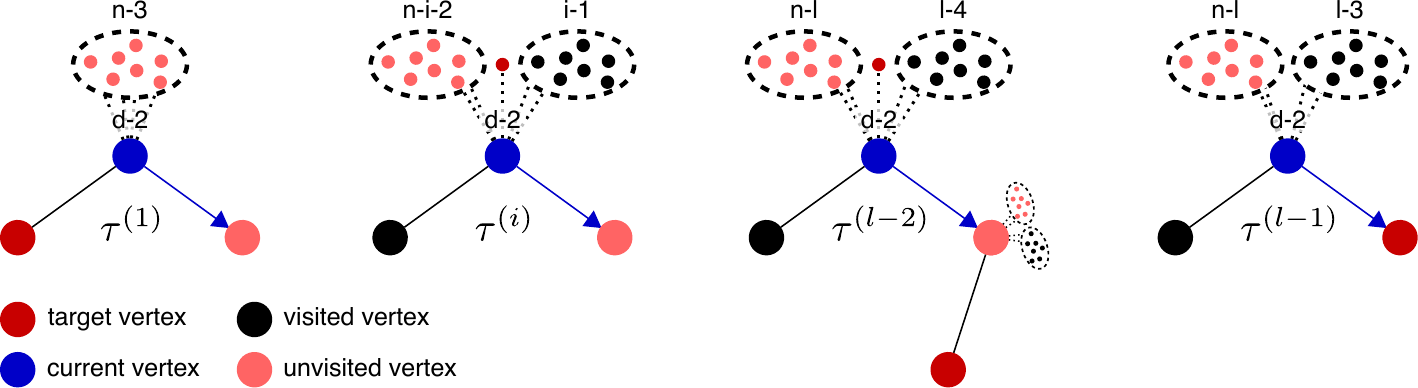}
  \caption{Mean-field treatment of the Laplacian-random-walk transitions.
  Red vertices have fixed potential one, blue vertices are current, and black
  vertices have already been visited.  The first and final two transitions
  receive separate corrections because adjacency to the target is known.
  Appendix~\ref{app:approximation} gives the formulas.}
  \label{fig:tau-approx}
\end{figure*}

\input{03-importance-sampling}

\input{03-approximation}

\input{06-theoretical}

\input{07-evaluation}

\input{04-cycle-counting}

\input{05-cycle-balance}

\input{08-conclusion}

\makeatletter
\@secpenalty=10000 
\makeatother

\section*{Code and Data Availability}
The implementation, experiment scripts, and graph-processing pipeline are
available in the
\href{https://github.com/josefhoppe/random-abstract-cell-complexes/}{project
repository}.  The cycle-sampling implementation is also distributed
as the \href{https://pypi.org/project/py-raccoon/}{\texttt{py-raccoon}}
package on the Python Package Index.  The
real-world networks used in the experiments are identified by their
Netzschleuder dataset names in the appendix figures.

\section*{AI usage declaration}
AI was used to assist with writing code for data visualization and handling of computational experiments.
All algorithmic functionality was implemented by hand, and all AI-generated code was carefully vetted.
The authors also used generative AI to assist with grammatical and language editing of the manuscript as well as revision of passages; and to find potential errors in the arguments and exposition.
All outputs and edits were critically reviewed, manually verified, and revised by the authors.
The authors claim full responsibility for the final version of the manuscript.

\begin{acknowledgments}
Funded by the European Union (ERC, HIGH-HOPeS, 101039827). Views and opinions expressed are however those of the author(s) only and do not necessarily reflect those of the European Union or the European Research Council Executive Agency. Neither the European Union nor the granting authority can be held responsible for them.
\end{acknowledgments}
\makeatletter
\@secpenalty=-300 
\makeatother

\bibliography{main}
\bibliographystyle{apsrev4-2}

\clearpage
\appendix
\crefalias{section}{appendix}

\input{99-appendix}

\end{document}

%% file: 00-abstract.tex
Simple cycles encode feedback and alternative routes in networks and are important for routing problems, the convergence of dynamical processes on networks, message passing and belief propagation, or even static network characterizations such as structural balance in signed networks.
Being able to assess and sample from the cycle structure of a graph is thus of interest in many analysis contexts.
However, due to the combinatorial growth of the cycle space with graph size, the computational cost for such sampling is high.
Therefore, we here introduce a sampling framework that enables us to sample from the cycle structure without enumerating all cycles.
Specifically, the method draws uniform spanning trees and collects the fundamental cycles created by their non-tree edges, such that each tree provides an informative subset of simple cycles.
Because cycles are included with unequal probabilities and are dependent within each subset, we formulate an inclusion-probability-based Horvitz--Thompson estimator, which we average over random subsets, a procedure we call \emph{Monte Carlo Horvitz--Thompson}.
For this, we sample multiple Horvitz--Thompson estimators and introduce a method to further subsample the individual estimators for increased computational performance.
For sampling cycles, the estimator is unbiased if exact cycle inclusion probabilities are known.
To extend the method to larger networks, we derive an efficient mean-field approximation for the inclusion probabilities and characterize the graph regimes in which it remains accurate.
Experiments on synthetic and real-world networks demonstrate the computational scaling and statistical performance of the framework.
We then illustrate the utility of Monte Carlo Horvitz--Thompson estimation by estimating cycle counts and cycle-level frustration statistics in signed spin systems.
An implementation is available in the \texttt{py-raccoon} package.

%% file: 01-introduction.tex
\section{Introduction}

Simple cycles, i.e., closed paths with no repeated vertices other than the initial and terminal vertex, are among the most basic nonlinear structures in a graph.  
They identify alternative routes, feedback, topological holes, and local redundancy, and their abundance across different lengths provides information that is not accounted for by degrees or pairwise distances alone.  
Cycle structure consequently appears in a broad range of network-analysis problems.  
In graph classification, cycle counts can provide features that increase the expressive
power of graph neural networks~\cite{bouritsas2022improving} beyond the Weisfeiler--Leman limit \cite{morris2019weisfeiler,xupowerful}.  
In signed social networks, the parity of negative edges around cycles is central to structural balance and polarization \cite{cartwright1956structural,giscard2017evaluating,talaga2023polarization}.
Cycles also describe circular exchange opportunities in financial networks \cite{cui2018circular} and can be used to reconstruct cyclic structure from observed flows \cite{hoppe2023representing}.
In cellular pathway analysis, elementary flux modes are directed cycles that form positive feedback loops \cite{zanghellini2013elementary,ullah2020towards} and are widely studied.
In ecological network analysis, the relevance of cycles compared to non-cyclical paths is described by different cycle-based metrics \cite{finn1976measures,han1997several,allesina2004cycling}.

However, computing simple-cycle statistics presents a computational bottleneck.
The number of simple cycles can grow exponentially with graph size, and exact counting is computationally hard in general~\cite{valiant1979complexity,harary1971number}.
Exact enumeration is therefore most practical for graphs with manageable numbers of cycles or analyses restricted to short cycles~\cite{harary1971number,alon1997finding}; enumerating longer cycles is typically expensive even with strong algorithmic optimizations~\cite{giscard2017evaluating,giscard2019general}.

This limitation is noted and dealt with differently in a wide variety of works.
For some applications, researchers limit their analysis to graphs where an exhaustive enumeration or exact computation is possible \cite{dominguez2014inherent,allesina2004cycling}.
In research on elementary flux modes, many different task-specific optimizations exist, but do not fully meet the need for scalability yet and research is ongoing \cite{ullah2020towards}.
Another alternative is to replace simple cycles by closed walks, which are more amenable to methods from linear algebra \cite{talaga2023polarization}.  
This substitution is always possible, but a closed walk may revisit vertices and edges and therefore need not represent the same structural object as a simple cycle.

\begin{figure*}
    \centering
    \includegraphics[width=.8\linewidth]{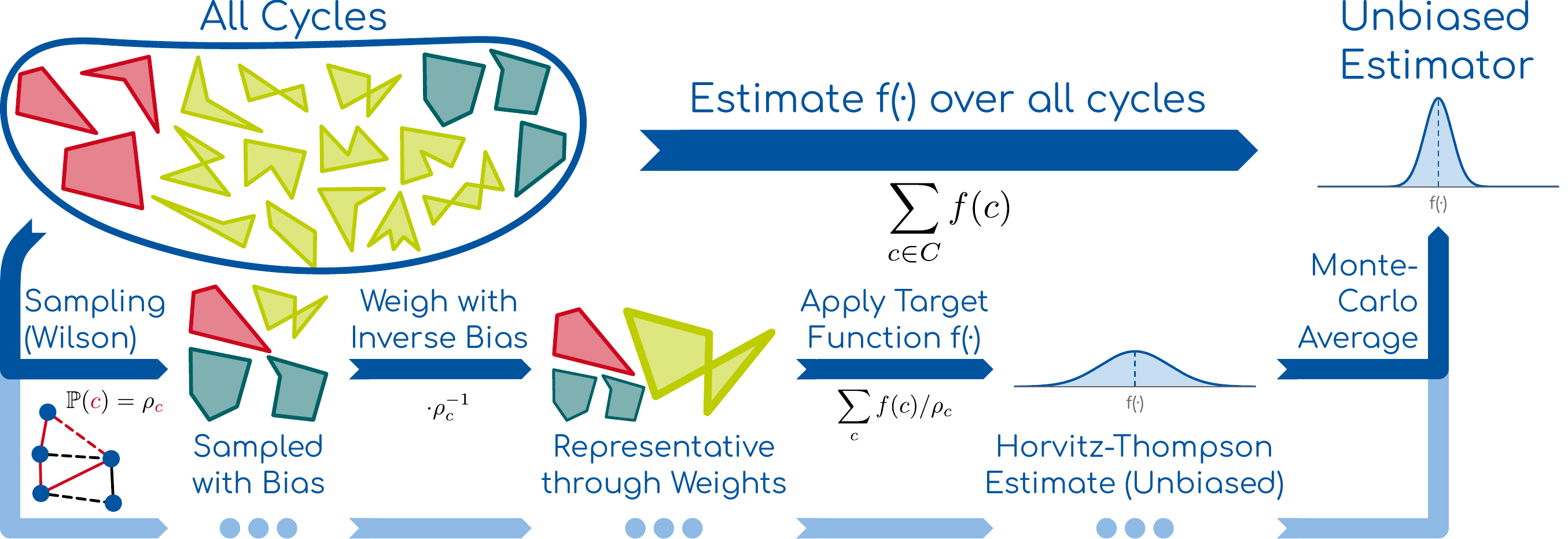}
    \caption{Overview of Monte--Carlo Horvitz--Thompson with spanning trees to estimate cycle statistics.}\label{fig:mcht}
\end{figure*}

Moreover, counting is only one of several relevant tasks associated with cycles.
Applications may require the computation of the distribution of a cycle statistic, an average over all cycles of a given length, or the fraction of cycles satisfying a property.  
Examples include the fraction of unbalanced cycles in a signed network, the distribution of gains over circular exchanges, and cycle-based anomaly or feedback scores.
These tasks motivate sampling methods that expose individual simple cycles without first enumerating the full population.
Ideally, we would draw cycles uniformly at random from all cycles of a chosen length.
In practice, direct uniform sampling is difficult because this population can be too large to enumerate and its size is generally unknown.

We address this problem by indirectly sampling simple cycles through their association with uniform spanning trees.
Every spanning tree provides a compact view of the graph's cycle structure: adding any non-tree edge to the tree creates one fundamental simple cycle.  
Sampling a uniform spanning tree therefore produces an entire subset of cycles in one iteration, and efficient algorithms such as Wilson's algorithm make the tree draw itself tractable.  
This construction has two important consequences.  
First, the cycles within a sampled subset are dependent, since they share the same tree.
Second, different cycles need not have the same probability of appearing: a cycle is included exactly when the sampled tree contains all but one of its edges.
Treating the observed cycles as uniform individual samples would consequently lead to biased estimates.

The appropriate correction for a single uniform spanning tree can be naturally accounted for in terms of inclusion probabilities via a Horvitz--Thompson~\cite{horvitz1952generalization} estimator.
It needs only the marginal probability that each object is included and does not require independence among objects within a subset or an explicit probability distribution over all possible subsets.  
Thus, cycles may share edges or vertices without compromising unbiasedness, although this dependence affects the variance.  
Using Monte Carlo averages over multiple such uniform spanning trees, we can then estimate the desired cycle statistics, closing the methodological gap we identified.

Here, the inclusion probabilities are determined by the spanning-tree sampling process, which contrasts with the standard usage of HT estimators, in which the sampling design can typically be chosen.
To further reduce the cost of evaluating these inclusion probabilities, we also consider an estimator that subsamples one cycle of the chosen length from each nonempty subset.
This additional sampling step preserves unbiasedness for totals when exact inclusion probabilities are used, with potentially higher variance than using the full subset.
We refer to Monte Carlo averaging of the per-tree HT estimates as \emph{Monte Carlo Horvitz--Thompson} (MCHT) estimation.

\paragraph*{Contributions.}
Our main contributions are as follows.
First, we develop Monte Carlo Horvitz--Thompson estimators for arbitrary object populations and apply them to compute cycle statistics as seen in \cref{fig:mcht}.
We clarify the relation of our estimators to importance sampling\footnote{For finite-population totals, ordinary importance sampling is closely related to Hansen--Hurwitz \cite{hansen1943theory} estimation. We use the importance-sampling terminology here.} and compare the full-subset estimator with one that selects one cycle from each nonempty sampled subset.

Second, we derive an exact expression for the inclusion probability of each fundamental cycle induced by a uniform spanning tree.

Third, we derive a mean-field approximation that permits near-linear evaluation of the inclusion probabilities for cycles of a fixed length induced by each sampled tree, in addition to the cost of generating the tree.
Using this approximation can introduce bias into the estimates.

Fourth, we characterize the limitations on sampled cycle lengths arising from uniform spanning tree diameters and empirically study runtime, probability-approximation error, and estimator convergence on synthetic and real-world graphs.

Finally, we demonstrate the framework for two tasks: estimating cycle counts and estimating frustration statistics in signed spin systems.

\paragraph*{Outline.}
\Cref{sec:bg-notation} introduces notation and preliminaries.
\Cref{sec:taming} describes the spanning-tree construction and exact inclusion probabilities, and \Cref{sec:subset-imp} develops Monte Carlo Horvitz--Thompson estimation.
\Cref{sec:approx} presents the scalable inclusion-probability approximation, \cref{sec:theoretical:limitations} discusses the lengths of cycles that can be sampled, and \cref{sec:experiments} evaluates runtime and the probability approximation.  
We then apply the estimators to cycle counting in \cref{sec:cycle-counting} and cycle-level frustration in \Cref{sec:frustration}.

%% file: 02-background.tex
\section{Background and Notation} \label{sec:bg-notation}
We use calligraphic letters for sets (e.g., $\mathcal X$), sans-serif uppercase letters for random objects (e.g., $\mathsf T$), and plain letters for their realizations (e.g., $T$).
For simplicity, given a set $\mathcal X$, we denote sampling from the uniform distribution over $\mathcal X$ as $x \sim \mathcal X$ instead of $x \sim \mathrm{Uniform}(\mathcal X)$.

\subsection{Graphs and Simple Cycles}
We consider undirected graphs $G=(\mathcal V,\mathcal E)$ with $n:=|\mathcal V|$ vertices and $m:=|\mathcal E|$ edges.
We denote the degree of a vertex $u$ by $d(u)$, the mean degree by $\bar d := \nicefrac{2m}{n}$, and the graph Laplacian by $L := D - A$, where $A$ is the adjacency matrix and $D = \mathrm{diag}((d(u))_u)$ the degree matrix of $G$.  
We write $\mathsf G\sim \mathcal{G}(n,p)$ for an \erdosrenyi{} (ER) random graph \cite{erdos1959random,gilbert1959random}: for each unordered pair of distinct vertices, the corresponding edge is present independently with probability $p$.  
In our mean-field calculations we will consider a graph $G$ as a realization of $\mathsf G$ to simplify calculations.

A cycle $c$ is a closed simple path on $G$; its length $|c|$ equals the number of vertices (or edges) in $c$.
We denote the set of all cycles in $G$ by $\mathcal C$.
We use $\mathcal C_l$ to denote the set of all cycles $c\in \mathcal C$ with length $|c|=l$ and $N_l = |\mathcal C_l|$ to denote the number of cycles of length $l$ on a graph.
If vertex $u$ or edge $(u,v)$ is part of the cycle $c$, we write $u \in c$ or $(u,v) \in c$, respectively.
Without loss of generality, we assume that all graphs are connected.
Indeed, on a graph with multiple connected components, cycles are contained within connected components, so counting and sampling can be performed independently on each component.

\subsection{Importance Sampling}
Importance sampling (IS) is a Monte Carlo technique to estimate expectations under a probability distribution $p$ when direct sampling from $p$ or direct evaluation of the expectation is difficult.
For a full introduction to importance sampling, we refer the reader to \cite{tokdar2010importance,mcbook,agapiou2017importance}.
Let $\mathsf X$ be a random element of $\mathcal X$ with probability distribution
$p$, that is, $\mathsf X\sim p$.
Given a function $f:\mathcal X \to \mathbb R$, we are now interested in estimating the quantity:
\[
\mu := \mathbb E_p[f(\mathsf X)],
\]
where the expectation is taken with respect to $p$.
The idea of importance sampling is to draw samples from an easier-to-sample proposal distribution $q_{\mathrm{prop}}$ and reweight the samples by
\[
w(x) := \frac{p(x)}{q_{\mathrm{prop}}(x)}.
\]
The ratio compensates for over- and undersampling: objects that are more common
under the proposal receive less weight, while rare objects receive more weight.
If $q_{\mathrm{prop}}(x)>0$ whenever $p(x)>0$ (support condition), then
\[
\mathbb E_p[f(\mathsf X)]
= \mathbb E_{q_{\mathrm{prop}}}\!\left[w(\mathsf Z)f(\mathsf Z)\right],
\qquad \mathsf Z\sim q_{\mathrm{prop}}.
\]
Thus, let us consider random objects $\mathsf Z_1,\dots,\mathsf Z_s
\stackrel{\mathrm{i.i.d.}}{\sim}q_{\mathrm{prop}}$ and denote their realized
values by $z_1,\dots,z_s$.  The importance-sampling estimate computed from
these realizations is
\begin{align}
    \hat\mu_{\mathrm{IS}}
    &= \frac{1}{s} \sum_{k=1}^{s} w(z_k) f(z_k)
    = \frac{1}{s} \sum_{k=1}^{s}
      \frac{p(z_k)}{q_{\mathrm{prop}}(z_k)} f(z_k).
\end{align}

In many applications, $p$ is known only up to a normalization constant (partition function), i.e.,
$p(x)=\tilde p(x)/Z$ with unknown $Z$.
Then we can use the self-normalized importance sampling estimator

\begin{align*}
    \hat\mu_{\mathrm{SNIS}}
    &= \frac{\sum_{k=1}^{s} \tilde w(z_k) f(z_k)}
            {\sum_{k=1}^{s} \tilde w(z_k)},\\
    \tilde w(x)
    &:=\frac{\tilde p(x)}{q_{\mathrm{prop}}(x)}
      =\frac{Zp(x)}{q_{\mathrm{prop}}(x)}.
\end{align*}
As the constant $Z$ cancels in the ratio, only unnormalized densities are needed.
In \cref{sec:subset-imp}, we use the symbol $q_{\mathrm{inc}}$ for inclusion probabilities under subset sampling; unlike $q_{\mathrm{prop}}$, $q_{\mathrm{inc}}$ is not a normalized probability mass function over cycles.

\subsection{Horvitz--Thompson Estimators}
Horvitz--Thompson (HT) \cite{horvitz1952generalization} (see also \cite{mackay2003information}) estimation is a method for estimating totals of a finite universe $\mathcal U$ where each sample consists of a random subset $\mathsf{S} \subseteq \mathcal U$.
The subset is characterized by unequal inclusion probabilities $\mathbb P(u \in \mathsf{S}) = \rho_u$ for $u \in \mathcal U$, which are not necessarily independent.
To unify notation with importance sampling and our method, we denote the inclusion probability by $\rho_u$ and represent the quantity to estimate as a function $f(\cdot)$.
The HT estimator now estimates the population total $Y = \sum_{u \in \mathcal U} f(u)$ for an element-wise quantity $f(u)$ from a concrete subset $S$ as follows:
\begin{equation}\label{eq:horvitz-thompson}
  \hat Y_{\mathrm{HT}}(S):=\sum_{u\in S}\frac{f(u)}{\rho_u}
\end{equation}

Like importance sampling, this estimator weights individual objects inversely to their inclusion probability.
For the same reason, this estimator is unbiased as long as $\rho_u > 0$ whenever $f(u) \neq 0$, i.e., $\mathbb E[\hat Y_{\mathrm{HT}}(\mathsf S)] = Y$.

Many applications use HT to design a single estimator, often for stratified sampling, or combine multiple different designs.
In contrast, we are considering a sampler of subsets that are defined by inclusion probabilities, which we cannot control.
Because we have no direct control of the inclusion probabilities, we adjust the effective sample size by combining multiple HT estimates into a Monte Carlo average instead.

There are also other methods in the literature that can be seen as HT sampling with identical inclusion probabilities.
For instance, the number of occurrences of treewidth-limited subgraphs can be estimated using random node colorings, considering only subgraphs that contain one node of each color \cite{alon2008biomolecular}.
Similarly, one can estimate the total number of triangles by only considering monochromatic edges and counting the triangles on these \cite{pagh2012colorful}.
Alternatively, edges can directly be removed with a certain probability, which leads to an efficient estimator for the number of triangles \cite{tsourakakis2009doulion}.
If applied or extended to cycles, all three approaches take exponential time in the length of the cycles to estimate the cycle count.

%% file: 03-algorithm.tex
\section{Taming the Cycle Space with Spanning Trees}\label{sec:taming}

The set of simple cycles in a general graph is difficult to characterize directly.
In this section, we use spanning trees to obtain a structured, but biased, view of that set.
By quantifying this bias, we can apply importance sampling to estimate cycle-level quantities.

\subsection{Cycle Space, Cycle Bases, and Uniform Spanning Trees}

Every spanning tree $T$ on a graph $G$ induces a set of (simple) cycles $\mathcal C_T$ (which is also a cycle basis \cite{syslo1979cycle}): each additional edge $(u,v)$ that is not part of $T$ creates a cycle consisting of the path from $u$ to $v$ on $T$ and the edge $(u,v)$.
The set of all such induced cycles is denoted by $\mathcal C_T$.
It contains exactly $m-n+1$ cycles, one for each non-tree edge.
For fixed cycle length $l$, we use $\mathcal C_{T,l} = \mathcal C_T \cap \mathcal C_l$.

A uniform spanning tree (UST) is a spanning tree picked uniformly at random from all spanning trees on a given graph.
Uniform spanning trees can be efficiently obtained, e.g., by Wilson's algorithm \cite{wilson1996generating}.
We denote the set of spanning trees by $\mathcal T(G)$, abbreviate it as
$\mathcal T$ when the graph is fixed, and write
$\mathsf T \sim \mathcal T$ for a random uniform spanning tree.
By Kirchhoff's matrix-tree theorem, the number of spanning trees in a graph
$G$ is the determinant of the graph Laplacian with one row and the
corresponding column removed \cite{CHAIKEN1978377}:
\begin{equation}\label{eq:treecount}
  \tau(G):=|\mathcal T(G)|=\det L^{(r)}.
\end{equation}
Here $L^{(r)}$ is the submatrix obtained by deleting the row and column of an arbitrarily chosen vertex $r$; the resulting determinant~\eqref{eq:treecount} does not depend on that choice.  
We adopt the standard convention $\tau(G)=1$ when $G$ consists of a single vertex, so that the determinant of the corresponding empty $0\times 0$ matrix is one.

Let $\mathcal P=(v_0,\ldots,v_k)$ denote a simple path, with vertex set
$\mathcal V(\mathcal P)$ and edge set $\mathcal E(\mathcal P)$.  The probability that all edges
of this path are included in a uniform spanning tree is
\[
\mathbb P\bigl(\mathcal E(\mathcal P)\subseteq \mathsf T\bigr)
=\frac{|\{T\in \mathcal T:\, \mathcal E(\mathcal P)\subseteq T\}|}{|\mathcal T|}.
\]
Besides certain random walk formulations \cite{lyklema1986laplacian,lawler1987loop}, this
path-inclusion probability can be computed by spanning-tree counting: if all
edges of the path $\mathcal P$ are contracted, then the spanning trees of the contracted graph are in bijection with the spanning trees of $G$ that contain
$\mathcal E(\mathcal P)$.
Hence,
\[
\mathbb P\bigl(\mathcal E(\mathcal P)\subseteq \mathsf T\bigr)
=\frac{\tau(G/\mathcal P)}{\tau(G)},
\]
where $G/\mathcal P$ denotes the graph obtained by contracting all edges of
$\mathcal P$, and both terms can be obtained from their respective Laplacian cofactors \eqref{eq:treecount}.

Note that as path contraction may create parallel edges (the contraction may be a multigraph), for the validity of the formula the multiplicities of the edges must be retained as edge weights in the contracted Laplacian.
As the Laplacian is invariant to the addition of self-loops, these loops do not contribute to the matrix-tree cofactor.

\subsection{Induced Cycle Inclusion Probabilities}

For our cycle sampling procedure, we need the inclusion probability $\rho_c=\mathbb P(c \in \mathcal C_{\mathsf{T}})$ of a cycle in the set of induced cycles of a tree, rather than the inclusion probability of a path.
How can we calculate this from the path inclusion probability?
First, we can write $\rho_c$ as a proportion of trees:
\[
\rho_c = \frac{\left|\{T\in\mathcal T: c \in \mathcal C_T\}\right|}{\left|\mathcal T\right|}
\]
Each cycle $c$ can be induced by any of the paths that are missing one edge $e$ in $c$:
\[
\rho_c = \frac{\left|\{T\in\mathcal T: c \in \mathcal C_T\}\right|}{\left|\mathcal T\right|} = \frac{\sum_{e \in c} \tau(G / (c \setminus \{e\}))}{\tau(G)}
\]

\begin{figure}
  \includegraphics[width=\linewidth]{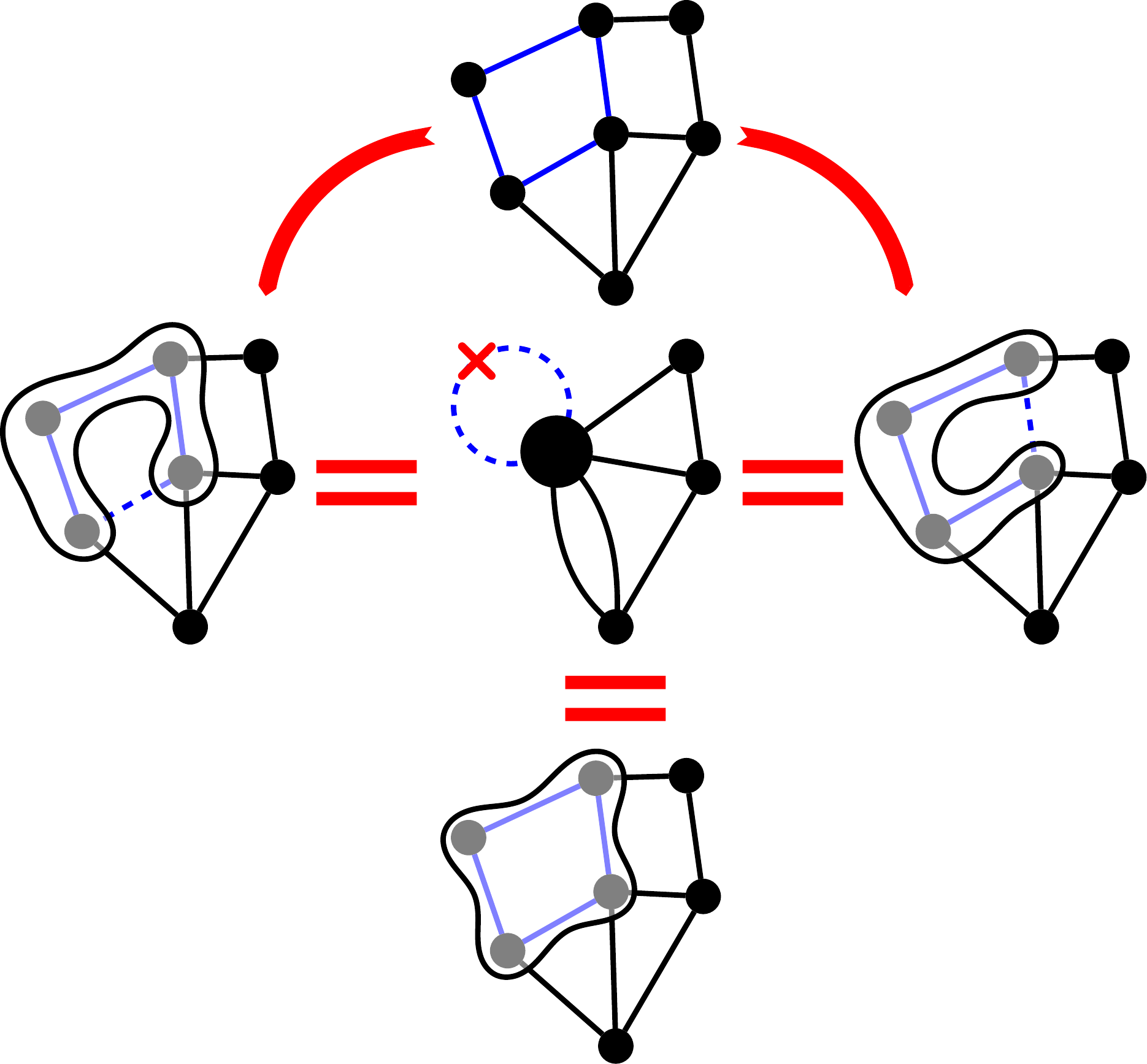}
  \caption{Top: A simple cycle on a toy graph. Middle: Two paths that, if included in a spanning tree, induce the cycle; and their edge contractions. Bottom: The cycle contraction. All three contractions result in the same multigraph (after deleting self-loops).}\label{fig:path-contraction}
\end{figure}

While we could compute this equation for $\rho_c$ in practice, we can further simplify it by looking at the different graphs $G / (c \setminus \{e\})$ we obtain by contracting the path $c \setminus \{e\}$ in $G$.
When contracting a path $\mathcal{P} \subset G$, all nodes in the path will be merged into one node and all edges between these nodes become self-loops.
\Cref{fig:path-contraction} illustrates this on a small toy graph.
The contracted node will thus have the same number of edges to every node (including itself), irrespective of the selection and order of edges that are contracted.
Thus, $G / c = G / (c \setminus \{e\}) \quad \forall e \in c$, i.e., the number of spanning trees only depends on the contracted nodes.
In other words, all paths have the same probability.
We simplify by multiplying the probability of any of the paths by the length of the cycle:
\begin{equation}\label{eq:exact-rho}
\rho_c = \frac{\sum_{e \in c} \tau(G / (c \setminus \{e\}))}{\tau(G)} = \frac{|c| \tau(G / c)}{\tau(G)}
\end{equation}

\subsection{Importance Sampling of Simple Cycles with Uniform Spanning Trees}

The construction of cycles via spanning trees also leads to a simple single-cycle sampling procedure for cycles of length $l$.
The idea is to first sample a tree $\mathsf T \sim \mathcal T$ and then a random non-tree edge, retrieving one cycle\footnote{Retrieving a random cycle from the set of induced cycles with the desired length is tempting, but since this set does not have a fixed size, the resulting inclusion probability is infeasible to compute.}.
The probability of obtaining this cycle is then
\begin{equation}\label{eq:is-prob}
  q_{\mathrm{prop}}(c) = \rho_c / |\mathcal C_T| = \rho_c / (m - n + 1)
\end{equation}
In practice, we often have to set $f(x)=0$ for cycles that do not have the correct length.
This is mathematically similar to rejecting the sampled cycle.
We could also explicitly implement rejection sampling by rescaling the inclusion probability $\rho_c$ by the inverse of the (empirical) probability to obtain a desired cycle.
For normalized expectations, this common factor cancels in a self-normalized
estimator.
For unnormalized totals such as a cycle count, it must instead be
known or estimated.

Depending on the graph, and especially in the limit for large graphs, this will lead to many estimates of $0$ or many rejections, respectively, leading to an increased variance of the estimator.
This makes it necessary to sample a large number of trees to obtain an accurate estimate, making the strategy impractical for many applications.
The core issue is the usage of only one cycle per tree in IS.
In the next section, we will introduce an alternative approach that samples multiple trees and averages over the Horvitz--Thompson estimators of the resulting cycle sets.

%% file: 03-importance-sampling.tex
\section{Monte Carlo Horvitz--Thompson with Uniform Spanning Trees}\label{sec:subset-imp}
The importance sampling construction in the previous section discards most of the cycles induced by a sampled tree.
In this section, we will discuss a per-tree HT estimator, using all induced cycles of the desired length.
Then, we derive an estimator more similar to standard importance sampling by subsampling the Horvitz--Thompson estimator.
This reduces the number of cycle inclusion probabilities we need to calculate, which speeds up computation at the cost of a higher variance in the estimator.
Finally, we show that both estimators have a substantially lower variance than the importance sampling approach introduced in the previous section.

The cycle sampling procedure directly lends itself to HT estimation:
Usually, we want to sample over the universe of cycles of a certain length $\mathcal U = \mathcal C_l$.
The random set $\mathsf S$ can be identified with the set of cycles $\mathcal C_{\mathsf T, l}$ from a uniform spanning tree $\mathsf T$, and we know the inclusion probability $\rho_c$ from \cref{eq:exact-rho}.
To get a more accurate estimate, we apply a Monte Carlo approach, i.e., we sample $s$ USTs $T_1, \ldots, T_s$ and average over the HT estimators:
\begin{equation}
  \hat{Y}_{\mathrm{MCHT}} = \frac{1}{s} \sum_{i=1}^{s} \hat{Y}_{\mathrm{HT}}(\mathcal C_{T_i,l}) = \frac{1}{s} \sum_{i=1}^{s} \sum_{c \in \mathcal C_{T_i,l}} \frac{f(c)}{\rho_c}
\end{equation}
The different spanning trees are independent, and the individual HT estimates are unbiased.
Thus, the Monte Carlo estimate is unbiased and its standard error is smaller than that of a single HT estimator by a factor of $\sqrt{s}$.

For a task estimating the average over a population (like in IS), we can use a self-normalized version.

The current estimator has one drawback: Unlike standard importance sampling, it requires the calculation of a large number of inclusion probabilities.
\Cref{sec:approx} introduces an approximation of the inclusion probability that can be efficiently calculated.
However, we can also use the Horvitz--Thompson estimator from above to derive a single-cycle estimator that is strictly better than the importance sampling introduced in \cref{sec:taming}.

For this, we use an unbiased sampler for the HT estimator:
\begin{equation}
  \mathbb{E}_{\mathsf{x} \sim \mathcal X}[|\mathcal X| \cdot f(\mathsf x)] = \hat{Y}(\mathcal X)
\end{equation}
Thus, the Monte Carlo one-cycle (MCOC) estimator is also unbiased (but may have a higher variance than Monte Carlo Horvitz--Thompson):
\begin{equation}
  \hat{Y}_{\mathrm{MCOC}} = \frac{1}{s} \sum_{i=1}^{s} |\mathcal C_{T_i, l}| \frac{f(c_i)}{\rho_{c_i}}, \quad c_i \sim \mathcal C_{T_i, l}, \mathsf T_i \sim \mathcal T
\end{equation}

We can compare this to IS as described in \cref{eq:is-prob}:
If at least one cycle of the desired length is induced by a tree $T$, the standard importance sampling approach returns one with probability
\[
p_{\mathrm{hit}} = \frac{|\mathcal C_{T,l}|}{|\mathcal C_T|}.
\]

\begin{algorithm}[tbp]
  \caption{MCHT cycle estimation from USTs}\label{alg:summary}
  \begin{algorithmic}[1]
\Require connected graph $G$, length $l$, tree count $s$, contribution
$f(\cdot)$, probability mode
\State $S\gets0$
\For{$i=1,\ldots,s$}
  \State sample a uniform spanning tree $T_i$
  \For{each non-tree edge $e=(u,v)$}
    \State let $c$ be $e$ plus the unique $u$--$v$ path in $T_i$
    \If{$|c|=l$}
      \State compute $a(c) = \rho_c$ or $\rho_c^{\text{MF}}$
      \State $S\gets S+f(c)/a(c)$
    \EndIf
  \EndFor
\EndFor
\State \Return $S/s$
\end{algorithmic}
\end{algorithm}

\begin{remark}[Effective sample sizes of MCOC and IS]
With growing graph size, we find longer cycles induced by almost all spanning trees.
For these cycles, MCOC provides one independent sample per sampled tree, whereas IS only provides a sample with probability $p_{\mathrm{hit}}$.
This smaller effective sample size translates to a larger variance in the estimator.
\end{remark}

This transitively implies that MCHT sampling leads to an even smaller variance.

\subsection{End-to-End Cycle Estimation}
\Cref{alg:summary} summarizes the overall procedure to estimate $\sum_{c \in \mathcal C_l} f(c)$ for one cycle length $l$.
In practice, we can estimate this total for all occurring lengths at the same time; the pseudocode only omits that for simplicity.

%% file: 03-approximation.tex
\section[Improved Scaling by Approximating the Cycle Inclusion Probability]{Improved Scaling by Approximating the Cycle Inclusion Probability}\label{sec:approx}

\begin{figure*}
    \centering
    \def\svgwidth{.7\textwidth}
    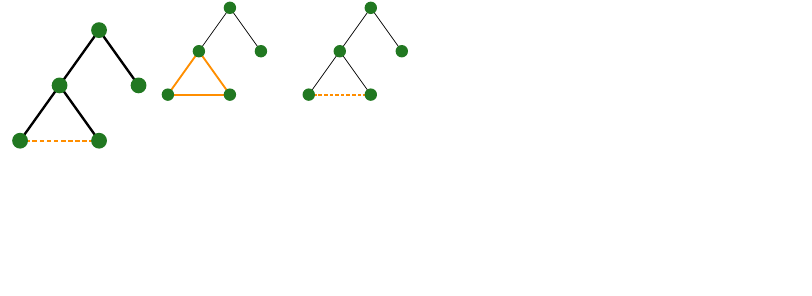
    \caption{Path aggregation on a rooted spanning tree.  Prefix sums and
    products from the root, together with lowest-common-ancestor queries, give
    the aggregates on every induced simple cycle.}
    \label{fig:treecalc}
\end{figure*}

The exact inclusion probability $\rho_c$ based on the determinant is useful for moderate problem sizes, but can become a computational bottleneck on larger graphs.
To avoid this, we design a mean-field approximation $\rho^\mathrm{MF}_c$ of $\rho_c$ that, using the sampled spanning tree, can be calculated in near-constant time per cycle.
The full derivation can be found in \cref{app:approximation}.
Here, we will only give an intuitive overview.

To achieve this, we replace the calculation of $\rho_c$ based on counting spanning trees by a calculation based on a loop-erased random walk, which is also how USTs are usually sampled \cite{wilson1996generating}.
The probability of a loop-erased random walk choosing a path is the same as the probability of a Laplacian random walk \cite{lyklema1986laplacian,lawler1987loop}.
The Laplacian random walk allows us to calculate step-wise transition probabilities.
Finally, we design a mean-field approximation of the individual steps that can be calculated efficiently over a spanning tree, i.e., decomposed into a multiplicative node-wise and additive edge-wise component.

A loop-erased random walk can be obtained by performing a random walk from a source $s$ until it hits a target $t$, erasing loops whenever a vertex is visited twice.
Wilson's algorithm \cite{wilson1996generating} samples uniform spanning trees by starting with a root node that is part of the tree, and then performing a loop-erased random walk from each vertex until it hits the tree, adding the walk to the tree.
Since USTs can be constructed this way, it is intuitive that loop-erased random walks between two points have the same probability distribution as taking the path between the same points from a UST (proven in \cite{pemantle1991choosing}).
However, the probability of obtaining a loop-erased random walk is infeasible to describe directly due to the loop erasure.

Instead, we use the Laplacian random walk, which shares the same probability distribution \cite{lyklema1986laplacian,lawler1987loop}.
Intuitively, the Laplacian random walk biases the transition, weighting it by the probability that a loop-erased random walk taking that transition will reach the target without erasing a loop.
Trivially, this means that the target node has weight $1$, and visited nodes (including the current node) have weight $0$.
All other nodes have a weight that is the average of their neighbors' weights (since the loop-erased random walk chooses from all neighbors uniformly).
The resulting path probability is exact and equivalent to the one obtained using the determinant.

To speed up computation, we apply a mean-field approximation, as illustrated in \cref{fig:tau-approx}.
This approximation can only use limited information in order to still fit the efficient calculation over the spanning tree.
First, we approximate the weight of all unvisited nodes with the mean-field approximation, i.e., based on the number of target nodes and visited nodes.
We can divide the steps into the first, last, second-to-last, and intermediate steps.
In the first step, one of our neighbors is the target node, and all other neighbors are unvisited.
In intermediate steps, we know that one neighbor (the next node) is unvisited and one neighbor (the previous node) is visited; for all remaining neighbors, we assume mean-field connectivity to all other nodes (visited, unvisited, target).
In the penultimate step, we have additional information about the next node, namely that it has the target node as a neighbor.
In the final step, the next node is the target node, and we assume mean-field connectivity to visited and unvisited nodes.

For $n>3$ and $l>3$, these factors can be combined into functions only depending on global properties and local node degrees as 
\begin{equation}\label{eq:approx-rho-final}
  \rho_c^{\mathrm{MF}}
  =\gamma(n,\bar d,l)
   \left(\sum_{(u,v)\in c}\eta(n,d(u),d(v))\right)
   \prod_{u\in c}\varrho(n,d(u))
\end{equation}

This mean-field approximation consists of a global correction factor $\gamma$, a per-edge factor $\eta$ that needs to be summed over the path, and a per-node factor $\varrho$ that needs to be multiplied over the path.
Instead of iterating over the induced cycles, we can decompose the sum and product components into paths from the root to different nodes.
The following subsection discusses how we can efficiently calculate the mean-field approximation.

\subsection{Algorithmic Complexity} \label{sec:theoretical:complexity}
After introducing our approximation $\rho^{\mathrm{MF}}_c$, we now analyze the computational complexity of calculating it for all cycles induced by one spanning tree.

\begin{theorem}
    Let $G$ be a connected graph with $n>3$, fix a length $l>3$, and let
    $W_G$ denote the expected number of
    random-walk steps required by one run of Wilson's algorithm for the chosen
    root and vertex order.  For $s$ independent trees, computing our
    cycle inclusion probability approximation for every induced cycle of
    length $l$ has expected runtime
    \[
       \mathcal O\!\left(s[W_G+(n+m)\alpha(n)]\right) \subset \mathcal{O}(s[W_G + m^{1+\epsilon}]).
    \]
\end{theorem}
Apart from the graph-dependent cost $W_G$ of drawing the tree, one sample
requires essentially linear work in the graph size; the inverse Ackermann
factor grows so slowly that it is below five for all practical input sizes.

\begin{proof}
Wilson's algorithm \cite{wilson1996generating} takes time linear in the number
of random-walk steps it performs; its expected contribution is therefore
$W_G$.  This quantity depends on the graph, root, and vertex order and should
not be identified with a single pairwise hitting time.  Hitting-time results
nevertheless explain why Wilson sampling is fast on many dense random graphs
and can be much slower on worst-case graphs
\cite{brightwell1990maximum,lowe2014hitting}.

For each occurring length $l$, the global factor $\gamma(n,\bar d,l)$ is computed once.
Each cycle $c \in \mathcal C_T$ is induced by a non-tree edge $(u,v) \in G \setminus T$ and consists of the edge $(u,v)$ and the path $\mathcal P_{u,v}$ from $u$ to $v$ in the tree.
We split the calculation of \cref{eq:approx-rho-final} into $\rho_c^{\mathrm{MF}}=\gamma(n,\bar d,l)[\eta(n,d(u),d(v))+\sigma(u,v)]\pi(u,v)$, where $\sigma(u,v)$ denotes a sum and $\pi(u,v)$ a product over the path $\mathcal P_{u,v}$.
Defining $\sigma(u,v)=\sum_{(x,y)\in \mathcal E(\mathcal P_{u,v})}\eta(n,d(x),d(y))$ and $\pi(u,v)=\prod_{x\in \mathcal P_{u,v}}\varrho(n,d(x))$ recovers $\rho_c^{\mathrm{MF}}$, because $c$ is the union of the non-tree edge $(u,v)$ and the tree path $\mathcal P_{u,v}$.

To calculate $\sigma(u,v)$ and $\pi(u,v)$ for the required vertex pairs, we
start at the root $r$ and traverse the tree (cf. \cref{fig:treecalc}).
From the definitions, $\pi(r,r)=\varrho(n,d(r))$ and $\sigma(r,r)=0$.
For a vertex $u$ with parent $v$, we calculate
$\pi(r,u)=\pi(r,v)\varrho(n,d(u))$ and
$\sigma(r,u)=\sigma(r,v)+\eta(n,d(u),d(v))$.
We can then obtain $\sigma(u,v)$ and $\pi(u,v)$ in constant time from the
lowest common ancestor of $u$ and $v$:

\begin{align}
\pi(u,v) &= \frac{\pi(r,u) \pi(r,v)
  \varrho(n,d(\lca(u,v)))}{\pi(r,\lca(u,v))^{2} }
  \label{eq:treecalc_pi}\\
    \sigma(u,v) &= \sigma(r,u)+\sigma(r,v)-2\sigma(r,\lca(u,v))
    \label{eq:treecalc_sigma}
\end{align}

We can calculate $\pi(r,u)$ and $\sigma(r,u)$ for all vertices $u$ in $\mathcal O(n) \subseteq \mathcal O(m)$ by traversing the tree from the root downwards.
\Cref{eq:treecalc_pi,eq:treecalc_sigma} have to be calculated only once for each edge that is not part of the tree, i.e., $\mathcal O(m)$.
To compute the lowest common ancestor (lca) of all required pairs, we use
Tarjan's off-line lca algorithm \cite{tarjan1979applications}, which runs in
$\mathcal O((n+m)\alpha(n))$ per tree, where $\alpha$ is the inverse Ackermann
function.

By multiplying with the number of sampled spanning trees $s$, the theorem follows.
\end{proof}

\begin{remark}[Complexity of calculating cycle lengths]
    The length of the cycle induced by $(u,v)$ is
    $\depth(u)+\depth(v)-2\depth(\lca(u,v))+1$.
    Calculating the depth of all vertices in a tree requires a traversal in $\mathcal O(n) \subseteq \mathcal O(m)$; for each cycle, the calculation takes constant time.
\end{remark}

\subsection{Limits of the Probability Approximation}\label{sec:theoretical-approx}

The approximate probability of sampling a cycle only depends on the length and the degrees of the vertices.
As an example, consider $k$-regular graphs, where all vertices have the same degree.
Therefore, the approximated inclusion probability cannot distinguish cycles of the same length and gives the same estimate for them.
However, different cycles may have different sampling probabilities.
The approximate probability fails to capture this.
In physical terms, the approximation retains local coordination numbers but
discards where those vertices sit in the global geometry.
Similar effects can happen on non-regular graphs, especially with strong structure.

In practice, this may or may not have a relevant impact depending on the metric considered.
If it is potentially relevant, the magnitude can be estimated by choosing a small number of cycles to calculate the probability exactly and comparing the probabilities to the approximate probability.
In \cref{sec:experiments}, we will empirically analyze the accuracy on different real-world and synthetic graphs.

%% file: tree-calculation.pdf_tex
\begingroup%
  \makeatletter%
  \providecommand\color[2][]{%
    \errmessage{(Inkscape) Color is used for the text in Inkscape, but the package 'color.sty' is not loaded}%
    \renewcommand\color[2][]{}%
  }%
  \providecommand\transparent[1]{%
    \errmessage{(Inkscape) Transparency is used (non-zero) for the text in Inkscape, but the package 'transparent.sty' is not loaded}%
    \renewcommand\transparent[1]{}%
  }%
  \providecommand\rotatebox[2]{#2}%
  \newcommand*\fsize{\dimexpr\f@size pt\relax}%
  \newcommand*\lineheight[1]{\fontsize{\fsize}{#1\fsize}\selectfont}%
  \ifx\svgwidth\undefined%
    \setlength{\unitlength}{389.41706139bp}%
    \ifx\svgscale\undefined%
      \relax%
    \else%
      \setlength{\unitlength}{\unitlength * \real{\svgscale}}%
    \fi%
  \else%
    \setlength{\unitlength}{\svgwidth}%
  \fi%
  \global\let\svgwidth\undefined%
  \global\let\svgscale\undefined%
  \makeatother%
  \begin{picture}(1,0.36485276)%
    \lineheight{1}%
    \setlength\tabcolsep{0pt}%
    \put(0,0){\includegraphics[width=\unitlength,page=1]{tree-calculation.pdf}}%
    \put(0.36224036,0.29327347){\color[rgb]{0,0,0}\makebox(0,0)[t]{\lineheight{1.25}\smash{\begin{tabular}[t]{c}$=$\end{tabular}}}}%
    \put(0.52650713,0.29326036){\color[rgb]{0,0,0}\makebox(0,0)[t]{\lineheight{1.25}\smash{\begin{tabular}[t]{c}$\cdot{}$\end{tabular}}}}%
    \put(0.68138012,0.29326036){\color[rgb]{0,0,0}\makebox(0,0)[t]{\lineheight{1.25}\smash{\begin{tabular}[t]{c}$\cdot{}$\end{tabular}}}}%
    \put(0,0){\includegraphics[width=\unitlength,page=2]{tree-calculation.pdf}}%
    \put(0.85529348,0.29170658){\color[rgb]{0,0,0}\makebox(0,0)[t]{\lineheight{1.25}\smash{\begin{tabular}[t]{c}$\div 2$\end{tabular}}}}%
    \put(0.12246307,0.34536204){\color[rgb]{0.01960784,0,0}\makebox(0,0)[t]{\lineheight{1.10000002}\smash{\begin{tabular}[t]{c}$r$\end{tabular}}}}%
    \put(0.00625846,0.27987954){\color[rgb]{0.01960784,0,0}\makebox(0,0)[t]{\lineheight{1.10000002}\smash{\begin{tabular}[t]{c}$lca(u,v)$\end{tabular}}}}%
    \put(0.02103146,0.15602514){\color[rgb]{0.01960784,0,0}\makebox(0,0)[t]{\lineheight{1.10000002}\smash{\begin{tabular}[t]{c}$u$\end{tabular}}}}%
    \put(0.12210064,0.15602514){\color[rgb]{0.01960784,0,0}\makebox(0,0)[t]{\lineheight{1.10000002}\smash{\begin{tabular}[t]{c}$v$\end{tabular}}}}%
    \put(0.26426735,0.18889529){\color[rgb]{0.01960784,0,0}\makebox(0,0)[t]{\lineheight{1.10000002}\smash{\begin{tabular}[t]{c}$\pi$\\$\sigma$\end{tabular}}}}%
    \put(0.4379278,0.18784244){\color[rgb]{0.01960784,0,0}\makebox(0,0)[t]{\lineheight{1.10000002}\smash{\begin{tabular}[t]{c}$\eta(n,d(u),d(v))$\\$\varrho(n,d(lca))$\end{tabular}}}}%
    \put(0.59280068,0.18889529){\color[rgb]{0.01960784,0,0}\makebox(0,0)[t]{\lineheight{1.10000002}\smash{\begin{tabular}[t]{c}$\pi(r,u)$\\$\sigma(r,u)$\end{tabular}}}}%
    \put(0.74767366,0.18889529){\color[rgb]{0.01960784,0,0}\makebox(0,0)[t]{\lineheight{1.10000002}\smash{\begin{tabular}[t]{c}$\pi(r,v)$\\$\sigma(r,v)$\end{tabular}}}}%
    \put(0.93502039,0.1888824){\color[rgb]{0.01960784,0,0}\makebox(0,0)[t]{\lineheight{1.10000002}\smash{\begin{tabular}[t]{c}$\pi(r,lca)$\\$\sigma(r,lca)$\end{tabular}}}}%
    \put(0,0){\includegraphics[width=\unitlength,page=3]{tree-calculation.pdf}}%
    \put(0.36224043,0.05271602){\color[rgb]{0,0,0}\makebox(0,0)[t]{\lineheight{1.25}\smash{\begin{tabular}[t]{c}$=$\end{tabular}}}}%
    \put(0.52650721,0.05270291){\color[rgb]{0,0,0}\makebox(0,0)[t]{\lineheight{1.25}\smash{\begin{tabular}[t]{c}$+$\end{tabular}}}}%
    \put(0.6813802,0.05270291){\color[rgb]{0,0,0}\makebox(0,0)[t]{\lineheight{1.25}\smash{\begin{tabular}[t]{c}$+$\end{tabular}}}}%
    \put(0,0){\includegraphics[width=\unitlength,page=4]{tree-calculation.pdf}}%
    \put(0.85529356,0.05114924){\color[rgb]{0,0,0}\makebox(0,0)[t]{\lineheight{1.25}\smash{\begin{tabular}[t]{c}$-2$\end{tabular}}}}%
  \end{picture}%
\endgroup%

%% file: 06-theoretical.tex
\section{On the Lengths of Sampled Cycles}\label{sec:theoretical:limitations}

Counting the number of cycles of an arbitrary length is \#P-hard \cite{valiant1979complexity}.
Like any approach, the spanning-tree-based method is limited by this, but it is interesting to see when and especially \emph{how} the limitation manifests.
In this section, we discuss the behavior of spanning trees and how this limits the length of efficiently sampleable cycles.

Trivially, every cycle on a graph can be induced by at least one spanning tree.
However, the length of induced cycles is at most one plus the diameter of the spanning tree.
The diameter of uniform spanning trees has been extensively studied.
For sufficiently high-dimensional graphs, the diameter is in $\Theta(\sqrt{n})$ \cite{michaeli2021diameter,aldous1990random,alon2022diameter}, including the complete graph $K_n$, dense \erdosrenyi{} graphs, and integer lattices of dimension $d > 4$.
In special cases, the expected diameter is larger.
For example, for 2D grid graphs, this is $\Theta(n^{\nicefrac{5}{8}})$ \cite{lawler2011conformal,beffara2008dimension}.

This does not make it impossible to obtain cycles of arbitrary length using spanning trees.
However, it requires sampling an ever-increasing number of trees to do so, limiting computational feasibility.

As a worst-case illustration, we consider Hamiltonian cycles (i.e., cycles of length $|c|=n$) on the complete graph $K_n$.
A Hamiltonian cycle is only induced by a Hamiltonian path (viewed as a spanning tree), of which $\nicefrac{n!}{2}$ exist in $K_n$.
Thus, the probability of finding a cycle $c$ of length $l=n$ is
\begin{align}
    \mathbb{P}(\mathcal C_{T,n} \neq \emptyset) &= \frac{n!}{2} \frac{1}{\tau(K_n)} = \frac{n!}{2 n^{n-2}} \leq \frac{1}{2} \frac{\lfloor n/2\rfloor!}{n^{\lfloor n/2\rfloor-2}}\\
    &\leq \frac{1}{2^{\lfloor n/2\rfloor - 2}} \quad \text{for $n$ sufficiently large}
\end{align}

In other words, to find cycles of length $n$, in the worst case we need to sample a number of spanning trees that is exponential in the size of the graph.
In practice, we see an exponential decrease in the probability of finding a cycle between $l \in \Theta(\sqrt{n})$ and $l = \Theta(n)$.

While there is no directly comparable approach in the literature, Giscard et al.~\cite{giscard2017evaluating} introduced an efficient cycle-based sampling algorithm that is still exponential in the length of sampled cycles.
Thus, our approach provides a relevant improvement in the length of cycles that can be efficiently sampled, even though the theoretical limitation makes the sampling of arbitrary cycles infeasible.

%% file: 07-evaluation.tex
\section{Empirical Evaluation} \label{sec:experiments}

Analogously to the theoretical considerations, we first evaluate the runtime of different approaches, then investigate the lengths of sampled cycles and the accuracy of the probability approximation.
Finally, we look at the accuracy and convergence in two application examples and use the results to discuss when to use which configuration.

The evaluation code is available at \repolink{}.

\begin{figure}[t]
    \vspace*{-.5cm}
    \centering
    \includegraphics[width=\linewidth]{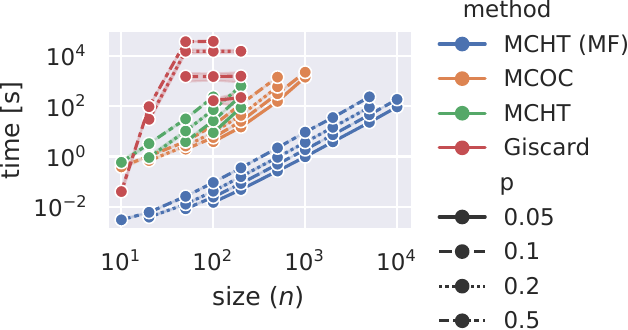}%
    \vspace{-4pt}
    \caption{Runtime on \erdosrenyi{} graphs for $s=100$ sampled trees and $10$ subgraphs of size $l_0=25$ (Giscard); mean and standard deviation over $10$ graphs ($5$ for Giscard).
    MCOC evaluates one exact inclusion probability for every length observed more than once per tree on average. Exact probability $\rho_c$ unless noted as MF.}
    \label{fig:time_exp}
\end{figure}

\begin{figure*}[ht]
    \begin{minipage}[b]{.47\linewidth}
        \centering
        \includegraphics[width=\linewidth]{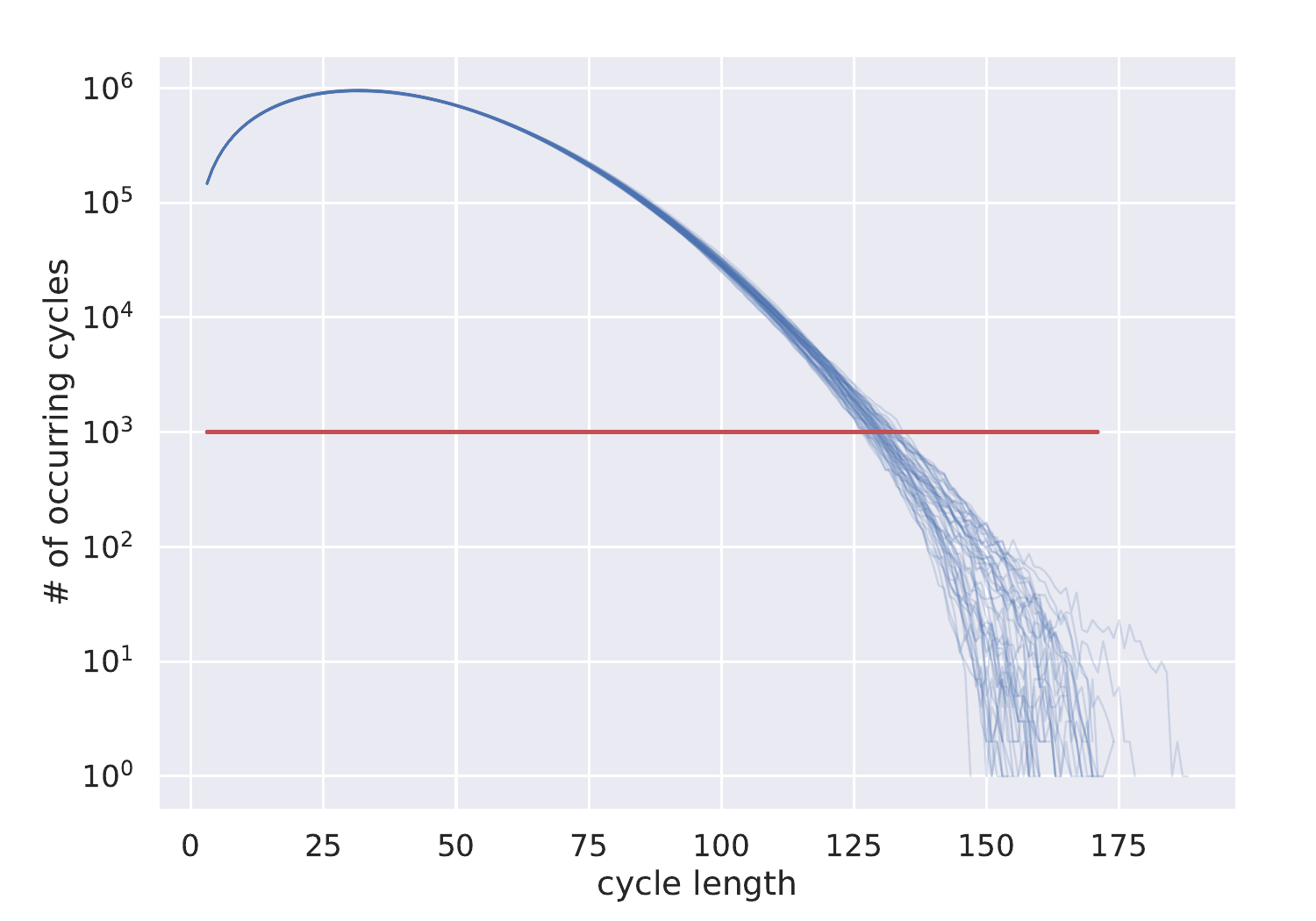}
        \panellabel{a}
    \end{minipage}
    \hfill
    \begin{minipage}[b]{.47\linewidth}
        \centering
        \includegraphics[width=\linewidth]{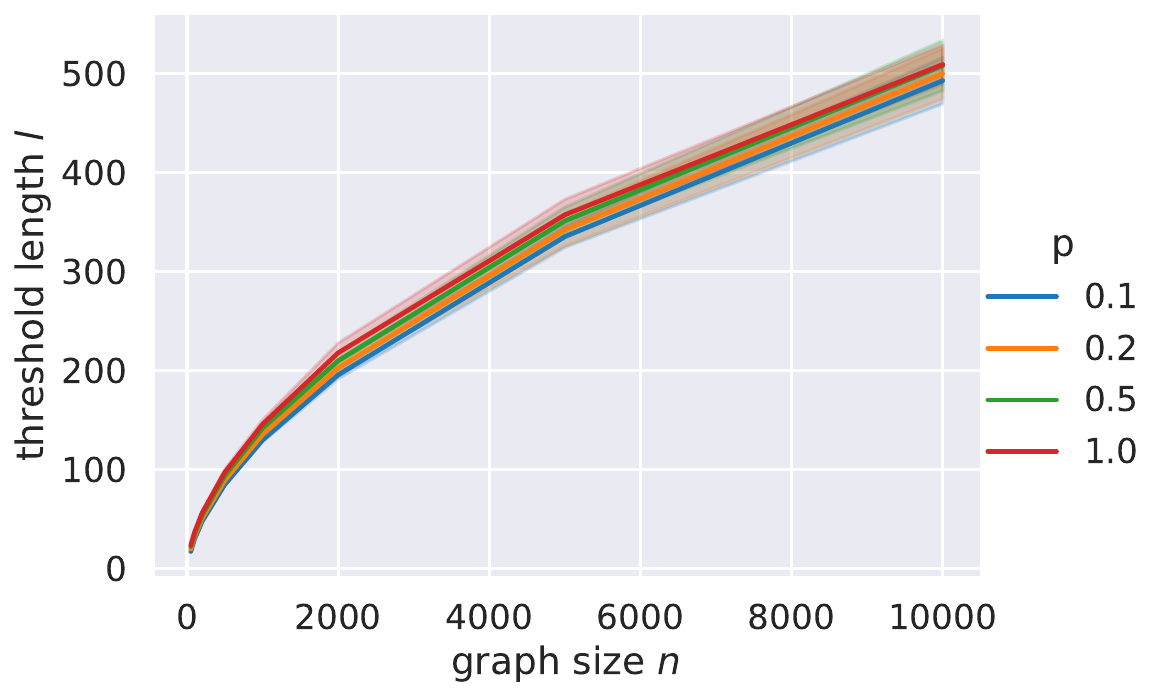}
        \panellabel{b}
    \end{minipage}
    \caption{Lengths of cycles exposed by uniform spanning trees.  (a) Number
    of induced cycles (by length) over $s=1000$ spanning trees on
    \erdosrenyi{} graphs with $n=1000, p=0.1$; one line for each of 50 runs on
    different graphs, the red line marks $s$ (one cycle per tree on average).  The most common length is near
    the $\sqrt n$ scale, while cycles with $l\approx130$ still appear about
    once per tree on average.  (b) Largest $l$ for which more than $1000$
    cycles were induced by $s=1000$ spanning trees; means and standard
    deviations are over 50 runs on different graphs for each $p$ and $n \in
    [50, 10\,000]$.  The largest
    well-observed length is empirically consistent with $\sqrt n$ scaling in
    this dense random-graph regime.}\label{fig:occ}
\end{figure*}

\subsection{Empirical Runtime}\label{sec:eval:time}

For the analysis of the runtime, we also compare to the current state-of-the-art Monte Carlo method for estimating balance by Giscard et al.\ \cite{giscard2017evaluating}.
The Giscard Monte Carlo method samples connected node-induced subgraphs of fixed size, uses an exact algorithm with exponential runtime to calculate the balance of the subgraph, and takes the average degree of balance as its overall estimate.

In their paper, Giscard et al.\ show measures for maximum cycle lengths between $15$ and $18$; they note that they obtained the best results for subgraph sizes $l_0$ between $20$ and $30$.
Thus, for our comparison, we also set $l_0=25$.
Note that with a constant number of sampled induced subgraphs and growing graph size, Giscard's algorithm cannot cover all nodes in the graph, whereas the spanning-tree-based approaches do.
Additonally, the spanning-tree-based approaches cover increasingly longer cycles.
In other words, the comparison is somewhat biased in favor of Giscard's algorithm.

\Cref{fig:time_exp} compares the runtime of the MCHT estimator with exact and approximated probabilities, the MCOC estimator with the exact probability, and the Giscard method.
Empirically, the runtimes for the approach are in line with our analysis as it takes approximately $\mathcal O(n^2)$, which is consistent with $\mathcal O(s\cdot{}m^{1+\epsilon})$ since $m \in \Theta(n^2)$ for a fixed $p$.
The MCHT estimator with approximated probabilities scales much better than any approach with exact probabilities, but the one-cycle estimator with exact probabilities is viable for larger graphs than MCHT with exact probabilities.
Giscard's estimator exhibits the longest runtime by a large margin.
We also evaluated the runtime on real-world networks (cf. \Cref{fig:time-rw}), which qualitatively lines up with the synthetic experiments.
Here, approximated MCHT is practical to use even for networks with more than $10^7$ edges.

All algorithms considered are highly parallelizable as each spanning tree or subgraph can be sampled independently (which is not currently implemented for the spanning-tree-based ones, but easy to achieve with the PyPI package).
Still, for practical applications, the size of the graph will determine which approaches are feasible, potentially excluding the slower ones altogether.

\subsection{Lengths of Efficiently Sampleable Cycles}

As mentioned in \cref{sec:theoretical:limitations}, for the dense random-graph regime considered here, the diameter of a uniform
spanning tree is of order $\sqrt n$.
\Cref{fig:occ} confirms the corresponding scale for the length of cycles that are frequently observed, and shows that, for ER graphs, this length grows from about $3\sqrt{n}$ at $n=50$ over $4\sqrt{n}$ at $n=1000$ to $5\sqrt{n}$ at $n=10\,000$, with only a small dependence on $p$.
With a decreasing number of cycles, MCHT and MCOC converge to the same estimator.

With increasing lengths, there is a steep decline in the number of cycles found.
In other words, finding cycles with a length of order of $\sqrt{n}$ is feasible, but finding longer cycles leads to an exponential number of samples required.

\subsection{Approximation Accuracy}\label{sec:eval:accuracy}

\begin{figure*}[t]
    \begin{minipage}{.48\linewidth}
        \centering
        \includegraphics[width=\linewidth]{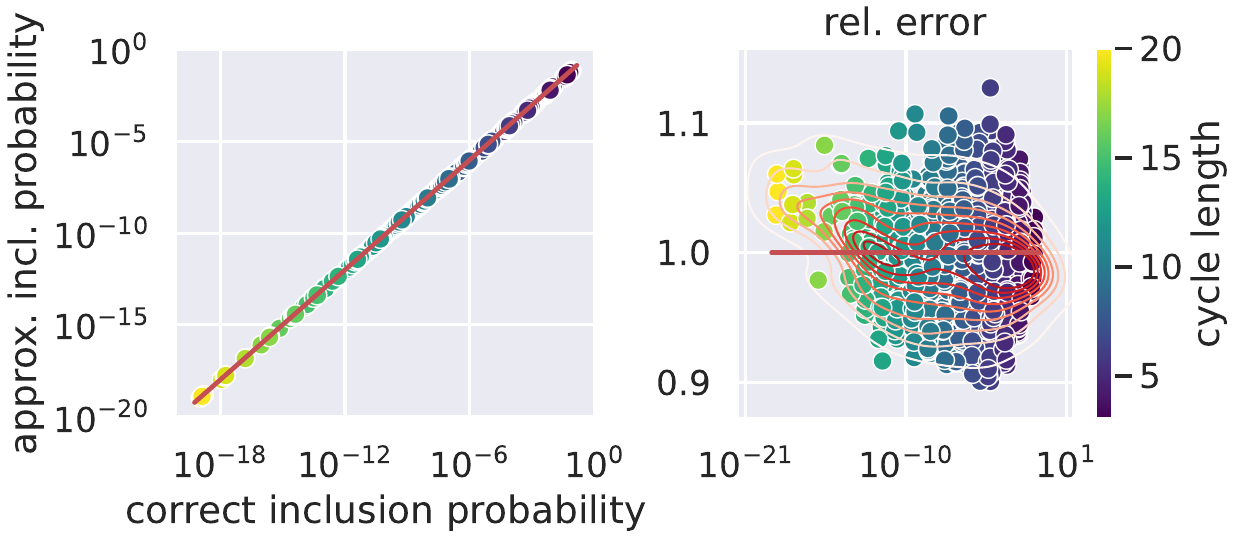}
        \panellabel{a}
    \end{minipage}
    \hfill
    \begin{minipage}{.48\linewidth}
        \centering
        \includegraphics[width=\linewidth]{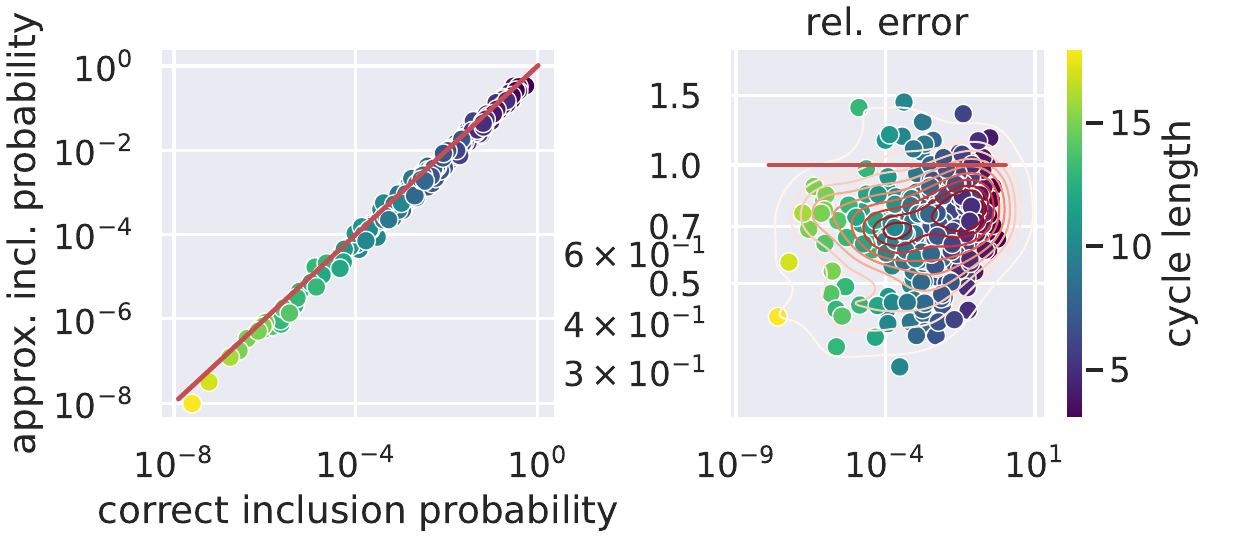}
        \panellabel{b}
    \end{minipage}
    \caption{Exact ($\rho_c$) versus mean-field ($\rho_c^{\mathrm{MF}}$) cycle
    inclusion probabilities for cycles exposed by ten USTs on Erd\H{o}s--R\'enyi
    graphs with $n=30$ and (a) $p=0.5$ ($209$ edges) or (b) $p=0.15$ ($63$
    edges); the right panels show the relative error
    $\rho_c^{\mathrm{MF}}/\rho_c$.  The approximation is more accurate on the denser
    \erdosrenyi{} graph.}
    \label{fig:estimation_both}
\end{figure*}

First, we compare the mean-field cycle inclusion probability
$\rho_c^{\mathrm{MF}}$ with the exact value on different \erdosrenyi{}
configurations.
Given a graph, we sample multiple spanning trees and calculate $\rho_c$ and
$\rho_c^{\mathrm{MF}}$ for every cycle $c$ induced by those trees.
Then, we repeat the procedure on other synthetic and real-world graphs to investigate the broader applicability of our approach to cycle sampling.
Because validation cycles are themselves obtained from uniform spanning trees,
these plots weight cycle $c$ in proportion to its inclusion probability
$\rho_c$.  They therefore describe approximation accuracy under the sampling
distribution, not uniformly over all cycles.  Rare cycles, whose inverse
weights can dominate estimator variance, are underrepresented.

\Cref{fig:estimation_both}(a) shows close agreement between exact and approximate probabilities in the denser graph.
A baseline using cycle length alone has visibly larger errors (\cref{fig:prob_mean_per_len}), showing that
local degree information improves the accuracy significantly.

However, as \cref{fig:estimation_both} (b) shows, for graphs that are relatively sparse, the accuracy drops.
This is likely a result of graphs with smaller edge probabilities behaving more differently from the expected graph:
Our approximation implicitly assumes a relatively densely and evenly connected graph.
Sparse configurations of ER are more likely to deviate from this expectation than dense configurations:
The degree distribution is less uniform and a (partial) path is more likely to `cut off' parts of the graph due to the lower connectivity.

For graphs that are not sampled from the \erdosrenyi{} model, the accuracy strongly depends on the structure.
If the graph is globally and relatively evenly connected, the approximation is still accurate.
We verified this with a number of synthetic examples in \cref{fig:appr_synth} (a)--(e), where our approximation is fairly accurate on most graphs.

On graphs with a macro structure that is uneven, the accuracy drops dramatically.
In a barbell graph (\cref{fig:appr_synth} (f)), all cycles are contained in one of the complete subgraphs.
However, the other vertices are still treated as unvisited and potentially
reachable, leading to an overestimate of the cycle inclusion
probability.\footnote{The mean-field model underestimates connectivity to the
target: within the clique the target is adjacent at every step, whereas the
approximation averages over possible target adjacency.}
The most notable deviations occur in Delaunay triangulations of the plane and two-dimensional
grid graphs (\cref{fig:appr_synth} (g) and (h)).
Here, we see a systematic underestimation of the cycle inclusion probability for long cycles.
This stems from the fact that the graph, while well-connected, has a large diameter and in nearly all steps, the target vertex is a neighbor with probability $0$ instead of $\nicefrac{1}{n}$.
More generally, this mechanism arises when different vertices are much easier or harder to reach, especially with a greater shortest path length (or diameter).
To a lesser degree, this can also be observed on the stochastic block model (SBM; \cref{fig:appr_synth} (c)), and likely many real-world networks.

Our experiments on real-world networks in \cref{fig:appr_realworld} support
these observations, with most graphs producing fairly small approximation
errors.
Graphs that have a comparatively large diameter or are (globally and nearly) planar show a systematic error, as indicated by our synthetic experiments.

%% file: 04-cycle-counting.tex
\section{Application: Approximate Cycle Counting}\label{sec:cycle-counting}

\begin{table*}
  \caption{Relative standard error of the mean ($\mathrm{SEM}/\mu$) and multiplicative bias (where statistically significant) of the cycle count estimation for $l=\lceil\sqrt{n}\rceil$, each sampled from $1000$ spanning trees. Note that the bias of all approximated approaches is the same in expectation, but due to differences in the variance, some biases are not statistically significant. `Count' derived from exact MCHT sampling. `er' is a $G(100, 0.1)$ graph, `grid' a $10 \times 10$ grid, and `delaunay' the Delaunay triangulation of $50$ uniformly random points. Real-world datasets retrieved from \textit{Netzschleuder}, \url{https://networks.skewed.de}, same naming convention.}\label{tbl:count-est}
\begin{tabular}{lrrlrrcrcrcrcrcrc}
\toprule
 &  &  &  &  & \multicolumn{4}{c}{MCHT} & \multicolumn{4}{c}{MCOC} & \multicolumn{4}{c}{IS} \\
\cmidrule(lr){6-9} \cmidrule(lr){10-13} \cmidrule(lr){14-17}
 &  &  &  &  & \multicolumn{2}{c}{exact} & \multicolumn{2}{c}{approx} & \multicolumn{2}{c}{exact} & \multicolumn{2}{c}{approx} & \multicolumn{2}{c}{exact} & \multicolumn{2}{c}{approx} \\
\cmidrule(lr){6-7} \cmidrule(lr){8-9} \cmidrule(lr){10-11} \cmidrule(lr){12-13} \cmidrule(lr){14-15} \cmidrule(lr){16-17}
Graph & $n$ & $l$ & Count & mean $|\mathcal C_{T,l}|$ & $\mathrm{SEM}/\mu$ & bias & $\mathrm{SEM}/\mu$ & bias & $\mathrm{SEM}/\mu$ & bias & $\mathrm{SEM}/\mu$ & bias & $\mathrm{SEM}/\mu$ & bias & $\mathrm{SEM}/\mu$ & bias \\
\midrule
polbooks & 105 & 11 & $1.53 \cdot 10^{9}$ & 14.8 & 0.03 & -- & 0.03 & 1.29 & 0.06 & -- & 0.06 & 1.26 & 0.29 & -- & 0.30 & -- \\
contact & 274 & 17 & $3.28 \cdot 10^{26}$ & 56.6 & 0.06 & -- & 0.06 & 1.75 & 0.09 & -- & 0.10 & 1.66 & 0.43 & -- & 0.44 & -- \\
grid & 100 & 10 & $1.58 \cdot 10^{3}$ & 6.0 & 0.01 & -- & 0.01 & 2.18 & 0.02 & -- & 0.02 & 2.18 & 0.12 & -- & 0.13 & 2.05 \\
er & 100 & 10 & $2.84 \cdot 10^{8}$ & 25.4 & 0.02 & -- & 0.02 & 1.20 & 0.03 & -- & 0.04 & 1.22 & 0.16 & -- & 0.17 & -- \\
delaunay & 50 & 8 & $5.06 \cdot 10^{3}$ & 3.8 & 0.02 & -- & 0.02 & 1.77 & 0.02 & -- & 0.03 & 1.78 & 0.16 & -- & 0.18 & -- \\
karate & 34 & 6 & $9.98 \cdot 10^{2}$ & 5.1 & 0.02 & -- & 0.03 & 2.08 & 0.03 & -- & 0.04 & 2.09 & 0.11 & -- & 0.14 & 2.16 \\
football & 115 & 11 & $1.38 \cdot 10^{9}$ & 20.0 & 0.01 & -- & 0.01 & 1.48 & 0.02 & -- & 0.02 & 1.50 & 0.19 & -- & 0.18 & -- \\
\bottomrule
\end{tabular}
\end{table*}

Counting the cycles of a given length is \#P-hard \cite{valiant1979complexity}.
However, we can easily design a Horvitz--Thompson estimator for the number of cycles of a certain length by setting $f(c) = 1$:
\begin{align}
  N_l&=|\mathcal C_l|=\sum_{c\in\mathcal C_l}1
  = \mathbb{E}\left[\sum_{c \in \mathcal C_{\mathsf T,l}} \frac{1}{\rho_c}\right], \mathsf T \sim \mathcal T
\end{align}
A cycle observed with probability $\rho_c$ represents, on average, $1/\rho_c$ cycles in the full population.
From this definition of $f$, we can directly derive MCHT and MCOC as unbiased estimators and substitute the mean-field approximation $\rho_c^{\mathrm{MF}}$ for a faster but potentially biased estimator.

This also works as an importance sampling estimator:
\begin{equation}
  \hat N_l^{\mathrm{IS}}
  := \frac{1}{s}\sum_{i=1}^{s}
     \frac{\mathbbm{1}\{|c_i| = l\}}{q_{\mathrm{prop}}(c_i)},
  \qquad c_i\stackrel{\mathrm{i.i.d.}}{\sim}q_{\mathrm{prop}}.
\end{equation}

\Cref{tbl:count-est} shows a comparison of results obtained using all of the aforementioned approaches in both exact and approximated variants.
As expected, the most accurate (lowest variance) approach is MCHT, followed by MCOC, and importance sampling is the least accurate by a large margin.
Using the approximated probability leads to a bias that depends on the graph's structure.
The error depends on the length of cycles considered, and is generally smaller for shorter cycles.

This error makes the approach a bad fit in some cases, e.g., where an unbiased estimate for the cycle count is required.
However, in many cases the error is actually tolerable for practical applications.
Different graphs often have cycle counts that vary by multiple orders of magnitude, far more than the systematic bias in many cases.
Furthermore, graphs with similar connectivity structures have similar biases.
Thus, a comparison or usage as an input feature for a downstream learning task may still work well even with this error.

In conclusion, for tasks such as cycle counting (where the end result directly relates to the bias in the probability approximation), it is sometimes required to use the exact version of an estimator.
The choice between MCHT and MCOC for the estimator is straightforward once we also consider the computational requirements for sampling a spanning tree and calculating the exact inclusion probability $\rho_c$.
The runtime of sampling a spanning tree is essentially negligible compared to calculating the exact inclusion probability $\rho_c$.
MCHT has a higher correlation between its sampled cycles, and thus MCOC with a bigger number of samples is preferable in the exact setting.
For the approximated version, MCHT and MCOC have practically the same runtime per tree, making MCHT a clearly preferable choice.

%% file: 05-cycle-balance.tex
\section{Application: Spin Glasses and Cycle Frustration}\label{sec:frustration}

\begin{figure*}[btp]
    \begin{minipage}{\linewidth}
        \includegraphics[width=\linewidth]{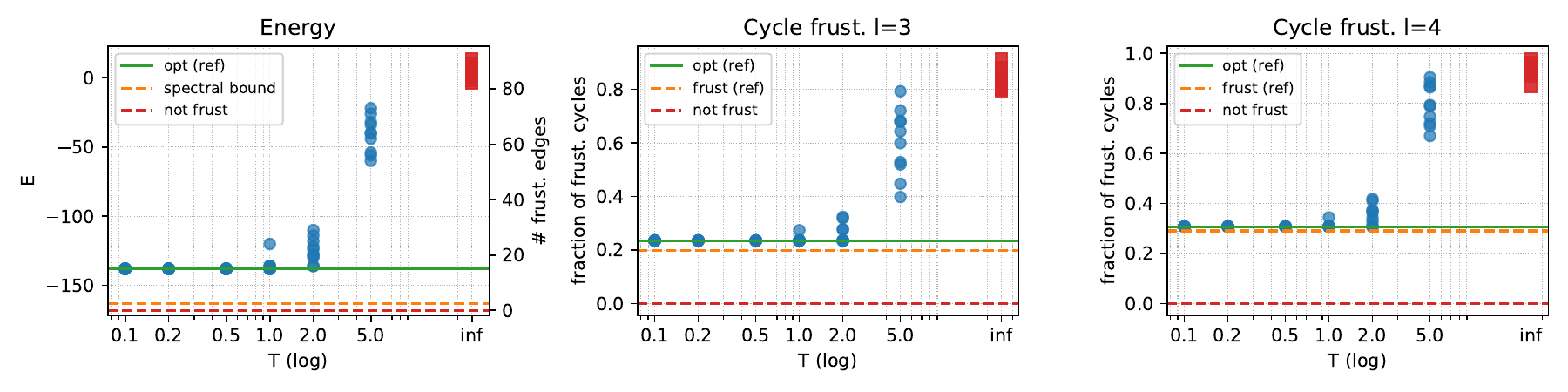}
        \panellabel{a}
    \end{minipage}
    \begin{minipage}{\linewidth}
        \includegraphics[width=\linewidth]{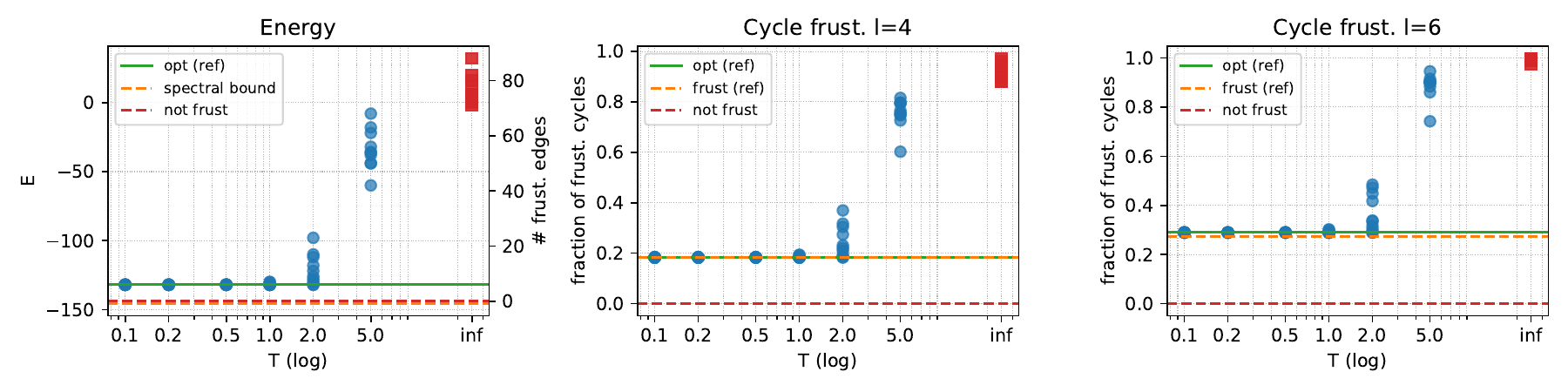}
        \panellabel{b}
    \end{minipage}
    \caption{Energy and cycle-level frustration for two spin systems: (a) an \erdosrenyi{} graph with $n=60$, $p=0.1$, and $p^-=0.1$ (for additional lengths, see \cref{fig:spin-er}) and (b) a $4\times4\times4$ grid graph with $p^-=0.05$ (for additional lengths, see \cref{fig:spin-grid}). Edges are selected to be negative couplings independently at random with probability $\mathbb P(J_{i,j} = -1) = p^-$. Cycle-level frustration is estimated with MCHT using the exact $\rho_c$ and $s=1000$ trees. Dots represent spin configurations $h$ obtained with temperature $T$ (x-axis), and red squares represent configurations obtained uniformly at random ($T=\infty$). The left plot shows energy and the fraction of frustrated edges; the other plots show the proportion of edge-frustrated cycles for selected lengths. Horizontal lines show values for a non-unique optimal configuration (green), the corresponding spectral or cycle-level lower bound (orange dashed), and zero frustration (red dashed). The energy and cycle-level quantities have different units and their bound tightness is not directly comparable.}\label{fig:spin-glass-good}
\end{figure*}

In this section, we consider how to apply cycle sampling to the area of spin glasses \cite{binder1986spin,toulouse1987theory} and investigate its performance.
A spin glass can be modeled as a graph where each vertex has a \emph{spin} value of $\pm1$, and two vertices can be positively or negatively coupled (i.e., have edges with value $\pm1$).
Here, positively coupled means that the coupling has a low-energy state when the two spins have the same orientation, and a high-energy state when they have opposite orientations.
Negative coupling is defined analogously.
The couplings are collected in a symmetric matrix $J$.
The Hamiltonian (energy function) of a state $h\in\{-1,1\}^{\mathcal V}$ is $E(h)=-\nicefrac{1}{2}h^\top J h$.
On the level of individual edges, we call an edge $(u,v)$ \emph{frustrated} if it is in the high-energy state, i.e., $J_{u,v} \cdot{} h_u \cdot h_v = -1$.
Equivalently, the spin assignment fails to satisfy the preferred relation encoded by that coupling.

For general interaction graphs, finding a ground state (a state with minimum energy) and its energy is NP-hard \cite{barahona1982computational}.
However, by simulating random changes to the spins, one can obtain a low-energy state.
Given a low-energy spin configuration obtained from this sampling process, it is interesting to see how close this spin configuration is to the ground state.

As an efficient approximation, there is a well-known spectral bound for the energy.
The largest eigenvalue $\lambda_{\max}$ of $J$ gives the lower bound
$E \geq - \frac{n}{2} \cdot \lambda_{\max}$.

As an alternative, we will now study frustration on the cycles.
We call a cycle \emph{edge-frustrated under $h$} if at least one of its edges is frustrated.
If there is an even number of negative couplings in the cycle, there is a spin configuration that leads to a non-frustrated cycle; for an odd number of negative couplings, such a configuration cannot exist.
Borrowing terminology from structural balance theory \cite{cartwright1956structural}, we call a cycle \emph{balanced} if it has an even number of negative couplings, and \emph{unbalanced} if it does not.
There is a spin configuration without frustrated edges if and only if every cycle is balanced.

The number of unbalanced cycles for a certain length is a lower bound for the number of edge-frustrated cycles (given any spin configuration $h$) of that length.
In some cases and for relatively short cycles, this cycle-level bound is quite tight. 
\Cref{fig:spin-glass-good} shows two cases where the cycle-level bound is tighter to the optimal count of balanced cycles than the spectral bound is to the energy.
For longer cycles, the balance bound becomes uninformative in all instances we investigated.
In other cases (cf. \cref{fig:spin-er2,fig:spin-grid2}), it is loose even at short lengths.

The number of unbalanced cycles is therefore a rigorous lower bound on the number of edge-frustrated cycles, but it is not by itself a bound on energy or on the ground-state optimality gap: one frustrated edge may belong to many cycles.
We consequently use it as a cycle-level diagnostic and do not compare its numerical tightness directly with the spectral energy bound.

To estimate the fraction of balanced cycles $b_G(l)$, we need to obtain an estimate for the number of balanced cycles and for the number of all cycles.
Trivially, the former can be achieved by setting $f^+(c)=1$ for balanced and $f^+(c)=0$ for unbalanced cycles:
\begin{align}
    \hat{b}_G(l) = \frac{\hat N_{l}^+}{\hat N_l} = \frac{\frac{1}{s} \sum_{i=1}^{s} \sum_{c \in \mathcal C_{T_i,l}} \frac{f^+(c)}{\rho_c}}{\frac{1}{s} \sum_{i=1}^{s} \sum_{c \in \mathcal C_{T_i,l}} \frac{1}{\rho_c}}, \mathsf T_1, \ldots, \mathsf T_s \sim \mathcal T
\end{align}

\subsection{Comparison of Approaches}

For this task, there is a previous Monte Carlo approach \cite{giscard2017evaluating}.
In their approach, Giscard et al.\ sample connected node-induced subgraphs and calculate the balance ratio on them.
As the authors note, this works as an estimator for the balance, but does not generalize to the cycle count.

Since the height of the spanning tree scales as $\Theta(\sqrt{n})$ and the number of induced cycles often drops off after $\sqrt{n}$, we show evaluation results for length $l=\left\lceil\sqrt{n}\right\rceil$.
Furthermore, we set the number of sampled subgraphs and spanning trees to $100$ and the size of subgraphs\footnote{Giscard et al.\ note that it is beneficial to set the maximum length bigger than the length that we are interested in and report using lengths between 20 and 30 while reporting results for lengths up to 18, but with increasing uncertainties.} (called $l_0$ in \cite{giscard2017evaluating}) to $\left\lceil2\sqrt{n}\strut\right\rceil$.

\begin{table*}
    \caption{Comparison of all balance estimation approaches on selected graphs, with $100$ samples (spanning trees / subgraphs of size $l_0=\lceil 2\sqrt{n}\rceil$, respectively). The ER graphs are $G(60, 0.1)$ (the same graph in the first two rows) and $G(100, 0.1)$; negative couplings are drawn independently with probability $p^- = 0.1, 0.3, 0.05, 0.05, 0.2, 0.08$ (top to bottom). Columns show the graph type, total size, number of positive and negative edges, balance as determined by MCHT-exact, and standard deviations over $100$ runs. The length shown is $l=\left\lceil\sqrt{n}\right\rceil$. All approaches have the same mean value for the balance (up to uncertainty). The entries highlighted in red indicate a statistically significant bias; all biases are smaller than $\pm 0.005$. Theoretically, MCOC and IS should have the same bias for the approximated variant, but it is too small to be statistically significant.}\label{tbl:balance-est}
\begin{tabular}{lrrrrrrrrrrrr}
\toprule
 &  &  &  &  &  & \multicolumn{2}{c}{MCHT} & \multicolumn{2}{c}{MCOC} & \multicolumn{2}{c}{IS} & \multicolumn{1}{c}{Giscard} \\
\cmidrule(lr){7-8} \cmidrule(lr){9-10} \cmidrule(lr){11-12} \cmidrule(lr){13-13}
Graph & $n$ & $l$ & $|\mathcal E^+|$ & $|\mathcal E^-|$ & $\hat{b}_G(l)$ & exact & approx & exact & approx & exact & approx &  \\
\midrule
ER & 60 & 8 & 152 & 16 & 0.56 & 0.02 & \textcolor{red}{0.02} & 0.07 & 0.08 & 0.23 & 0.23 & 0.06 \\
ER & 60 & 8 & 119 & 49 & 0.50 & 0.02 & 0.02 & 0.07 & 0.07 & 0.22 & 0.23 & 0.06 \\
ER & 100 & 10 & 453 & 21 & 0.70 & 0.02 & \textcolor{red}{0.02} & 0.06 & 0.06 & 0.25 & 0.26 & 0.04 \\
$4 \times 4 \times 4$ Grid & 64 & 8 & 138 & 6 & 0.68 & 0.02 & 0.02 & 0.05 & 0.05 & 0.16 & 0.18 & 0.09 \\
$4 \times 4 \times 4$ Grid & 64 & 8 & 113 & 31 & 0.49 & 0.02 & \textcolor{red}{0.02} & 0.06 & 0.07 & 0.19 & 0.21 & 0.09 \\
$10 \times 10$ Grid & 100 & 10 & 169 & 11 & 0.62 & 0.03 & 0.03 & 0.05 & 0.06 & 0.22 & 0.23 & 0.12 \\
\bottomrule
\end{tabular}
\end{table*}

\Cref{tbl:balance-est} shows the results.
We observe much smaller biases in the approximated versions of the estimators, to the point that the bias is only statistically significant for MCHT and only on half of the tested graphs.
In practical applications, this bias is likely negligible.
Since the estimator can only be biased by a correlation between the approximation error and $f^+$, the average approximation error is not relevant here; and since the variance in the approximation error is comparatively low, a negligible bias is generally very likely.
Giscard's algorithm is not only the slowest in our experiments, but also less accurate than MCHT and comparable to MCOC.
Standard importance sampling is, again, the least accurate approach.

In practice, this means that the accuracy of MCHT-approx is almost as good as that of MCHT-exact with the same number of trees.
However, since MCHT-approx is much more computationally efficient, it provides a better cost/accuracy trade-off in these cases.
This confirms that the selected estimation task is highly relevant for the selection of the most practically useful estimator.

%% file: 08-conclusion.tex
\section{Conclusion}

In this paper, we have introduced Monte Carlo methods based on uniform spanning trees and Horvitz--Thompson estimators to estimate properties of the cycle space on graphs from cycles induced by spanning trees.
While inherently limited by the theoretical complexity of the problem, our methods push the boundaries of computational feasibility to cycles that have a length on the order of $\sqrt{n}$, which is considerably further than the current state of the art that was exponential in the length of the cycles considered.
Furthermore, our approach is more general and may be used for a wide variety of cycle-based metrics and features.

For this, we need to adjust the estimator using the inclusion probability of individual cycles.
The exact estimators we introduced are already more computationally efficient than any current approach, but for maximum scalability, we have introduced a degree-based approximation of the inclusion probability such that the runtime of the overall estimator is almost linear in the number of edges.

The accuracy and runtime trade-off needs to be considered based on the estimation target and the underlying graph structure.
On some graphs, there is a significant approximation bias that increases for longer cycles and leads to inaccuracies for non-normalized estimates (e.g., cycle counting).
While the error may still be acceptable for some applications, it can make the approximated version infeasible for others, and usually using the exact one-cycle estimator leads to the best results in these cases.
On the other hand, when considering normalized estimates (e.g., the fraction of cycles fulfilling a certain property), this systematic bias is not a problem.
Theoretically, a correlation between the variance in the approximation and the estimation target could still lead to a bias in the overall result.
However, since this variance is much smaller than the bias in all graphs that we investigated, it is unlikely to lead to a large error in practice even if it is strongly correlated.
Thus, in these cases, it is usually preferable to use the full MCHT estimator with approximated probabilities.

Most importantly, this groundwork closes a gap in the methodological toolkit for multiple existing applications in different subject areas, enabling the scalability of unbiased analysis to larger networks.
In addition to enabling or extending existing methods, analogues to other global properties on the graph may be defined over the cycles.
Thus, there are applications in many fields, such as computational social science, biology, economics, and physics.

Beyond applying and analyzing the practical performance of the method, future work could further optimize the method.
Currently, we either calculate a full Horvitz--Thompson estimate or subsample it with a single cycle.
Subsampling a set number per length may lead to comparable variances, without the full computational cost.

%% file: 99-appendix.tex
\FloatBarrier

\onecolumngrid

\section[Approximation of the 
Cycle Inclusion Probability rho_c]{Approximation of the Cycle Inclusion Probability $\rho_c$}
\label{app:approximation}

A cycle is induced by a spanning tree if and only if all but one of its edges are part of the spanning tree.
Each path formed by removing an edge from the cycle has a certain probability of being part of the spanning tree.
By summing these probabilities, we can obtain the overall cycle inclusion probability $\rho_c$.

To approximate this, we briefly outline the loop-erased random walk that is
also used to sample uniform spanning trees.
We then present the Laplacian random walk
\cite{lyklema1986laplacian}, an equivalent formulation with the same path
distribution \cite{lawler1987loop}.
This gives an exact equation for the cycle inclusion probability $\rho_c$.
Finally, we introduce an approximation of that equation that can be calculated efficiently on spanning trees.

\subsection{The Loop-Erased Random Walk}

On a graph $G$, a loop-erased random walk from vertex $u$ to vertex $v$ is
obtained by performing a random walk and chronologically erasing loops whenever
the walk self-intersects.
Wilson's algorithm \cite{wilson1996generating} constructs a uniform spanning
tree from a sequence of such walks.
The loop-erased-walk path from $u$ to $v$ has the same distribution as the
unique $u$--$v$ path in a uniformly sampled spanning tree
\cite{pemantle1991choosing}.
This also implies that the loop-erased random walk from $u$ to $v$ has the same distribution as performing the loop-erased random walk from $v$ to $u$.

\subsection{The Laplacian Random Walk}

The Laplacian random walk (LRW) \cite{lyklema1986laplacian} from $u$ to $v$ has the same path distribution as the loop-erased random walk from $u$ to $v$.
The LRW achieves the same probability distribution as the loop-erased random walk by avoiding loops in the first place.
To do this, the transition probabilities are adjusted to account for the
possibility of creating a loop before reaching the target vertex $v$.
Previously visited vertices always lead to a loop, and are thus never chosen.
Choosing the target $v$ terminates the walk and can thus never create a loop.
For all other vertices, the probability of closing a loop depends on the following steps.
Since the loop-erased random walk chooses neighbors uniformly, their probability is the average of the probabilities of their neighbors.

Formally, in step $i$, each vertex $w$ is assigned a weight $f_i(w)$ depending on its relation to the random walk.
The probability to transition from $u$ to $w$ is then $\tau^{(i)}(w) = f_i(w) / \sum_{x \in \neigh(u)} f_i(x)$.
For the target, $f_i(v) = 1$.
For visited nodes (including the current one) $w$, $f_i(w) = 0$.
The weight of every other vertex $w$ is \emph{harmonic} in the Laplacian $L$, i.e., it satisfies $(Lf_i)(w) = 0$.
Since the set of visited nodes changes after each step, this weight has to be recalculated after each step and each transition probability $\tau_i$ depends on the previous transitions.

\subsection{Calculating the Cycle Inclusion Probability and Approximating the Laplacian Random Walk}

As discussed, a spanning tree induces a cycle $c$ if and only if all but one edge of $c$ belong to the tree.
Thus, each edge $e \in c$ corresponds to a potential path in a spanning tree.
For each edge $e = (v_0^{(e)},v_{l-1}^{(e)}) \in c$, choose an orientation\footnote{Since the probability of the loop-erased random walk does not depend on the direction, neither does the LRW, making the orientation arbitrary.} $\mathcal P_e=(v_0^{(e)},\ldots,v_{l-1}^{(e)})$ of the open path $c\setminus\{e\}$, where $l=|c|$.
At step $i$, let $f_{e,i}$ be the harmonic potential with target $v_{l-1}^{(e)}$ and visited set $\{v_0^{(e)},\ldots,v_i^{(e)}\}$.
With the previous definition,
\[
  \tau_{e,i} =\frac{f_{e,i}(v_{i+1}^{(e)})}{\sum_{y\in\neigh(v_i^{(e)})}f_{e,i}(y)}.
\]
The exact cycle inclusion probability is then
\begin{equation}
  \rho_c
  =\sum_{e\in \mathcal E(c)}\prod_{i=0}^{l-2}\tau_{e,i}.
\end{equation}

Since the LRW requires calculating $f_i$ for every step of the walk, it is too computationally expensive for our purpose.
Instead, we derive a mean-field approximation that assumes equal (weighted) connectivity for most edges in the graph but integrates local connectivity where possible.
To enable efficient computation over a spanning tree, we can only consider exact degrees on one edge (two adjacent vertices) in the path at the same time.
Additionally, depending on the step, we integrate structural information from the complete path.

The mean-field approximation assumes that unobserved nodes have mean degree $\bar d$ and are connected to all other nodes with weight $\frac{\bar d}{n-1}$.
For an \erdosrenyi{} graph, this is equivalent to the weights in the expected graph.
All vertices that have not been visited and are not the target are symmetric, and their potential only depends on the number of visited nodes.
This gives the mean-field estimate
\begin{equation}
  \forall u: f_i(u) \approx \begin{cases}
    1 & \text{if } u \text{ is the target}\\
    0 & \text{if } u \text{ has been visited}\\
    \widetilde f_i := \frac{i\cdot 0 + 1\cdot 1}{i+1}
      = \frac{1}{i+1} & \text{otherwise}
  \end{cases} 
\end{equation}

However, this first approximation ignores the local structure of the LRW.
For vertices adjacent to the current vertex, we can give a more accurate approximation.
The current vertex is adjacent and, by definition, has value zero, reducing the mean-field value of $f_i$.
While this is symmetric for unvisited neighbors, the target may also be a neighbor.
Thus, the transition probability to the prescribed next vertex is smaller than the global approximation $\widetilde f_i$ alone would suggest.
Instead, we can derive a mean-field approximation specifically for neighbors $f_i(x)\approx\widetilde f'_i$:
\begin{align}
\widetilde{f}'_i &\approx \frac{\overbrace{0 \cdot{} 1}^{\text{current vertex}} + \overbrace{\bar d - 1}^{\text{other neighbors}}\frac{\overbrace{0 \cdot{} (i-1)}^{\text{\scriptsize other visited}} + \overbrace{1 \cdot 1}^{\text{\scriptsize target } v} + \overbrace{\widetilde{f}_i \cdot (n-i-2)}^{\text{\scriptsize unvisited}}}{n-2}}{\bar d}\\
&= \frac{\bar d-1}{\bar d} \frac{1 + \frac{n-i-2}{i+1}}{n-2}\\
&= \frac{\bar d-1}{\bar d} \frac{n-1}{n-2} \frac{1}{i+1}.
\end{align}

Furthermore, for the penultimate vertex, one of the neighbors is the target
vertex $v$.
Thus, its $f_{l-2}$ is bigger, increasing the probability for this step.
We denote the corrected value by $\widetilde f''_{l-2}$:

\begin{align}
    \widetilde{f}_{l-2}'' &= \frac{\overbrace{0 \cdot{} 1}^{\text{current vertex}} + \overbrace{1 \cdot 1}^{\text{target } v} + \overbrace{\bar d - 2}^{\text{other neighbors}}\frac{\overbrace{0 \cdot{} (l-3)}^{\text{other visited}} + \overbrace{\widetilde{f}_{l-2} \cdot (n-l)}^{\text{unvisited}}}{n-3}}{\bar d}\\
    &= \frac{1 + \frac{\bar d - 2}{n-3}\frac{n-l}{l-1}}{\bar d}.
\end{align}

With these considerations, we approximate the transition probabilities of
individual LRW steps.
For the first step, no other vertex has been visited and all neighbors are
either unvisited or $v$.
The first-step approximation depends on $n$, $\bar d$, and the degree $d$ of
the initial vertex.
The target vertex $v$ is a neighbor because the edge $(u,v)$ closes the cycle.
The remaining $d-1$ neighbors, including the next vertex, are unvisited.

\begin{align}
  \widetilde\tau_1(n,d,\bar d)
  &= \frac{\widetilde f'_1}{1 + (d - 1)\widetilde f'_1} \\
  &= \frac{1}{(\widetilde f'_1)^{-1} + d - 1}\\
  &= \frac{1}{2\frac{\bar d}{\bar d - 1}
       \frac{n-2}{n-1} + d - 1}.
\end{align}

For an intermediate step $2 \leq i < l-2$, we can obtain a similar expression.
One neighbor was visited in the previous step, so only the remaining $d-1$ neighbors must be considered; one is the prescribed next vertex.
Among the $n-3$ possible vertices in the mean-field calculation, $i-2$ have already been visited, one is the target $v$, and $n-i-2$ are unvisited.
The corresponding fractions therefore have denominator $n-3$.

\begin{align}
\widetilde\tau_i(n,d)
&= \frac{\widetilde f'_i}
        {\widetilde f'_i
        +(d-2)\frac{(n-i-2)\widetilde f'_i+1}{n-3}} \\
&= \frac{1}{1+\frac{d-2}{n-3}
        \left(n-i-2+(\widetilde f'_i)^{-1}\right)}.
\end{align}

However, this expression depends on the current step $i$, which would make it impossible to compute from path-wise aggregates over spanning trees.
Instead, we will approximate intermediate steps further by using $\widetilde{f}_i' \approx \widetilde{f}_i$:
\begin{align}
\widetilde \tau'_i(n,d)
&:=\frac{\widetilde f_i}{\widetilde f_i+(d-2)\frac{1}{n-3}((n-i-2)\widetilde f_i + 1)} \\
&=\frac{1}{1+(d-2)\frac{1}{n-3}\left(n-i-2+\widetilde f_i^{-1}\right)}
 =\frac{1}{1+(d-2)\frac{n-1}{n-3}}.
\end{align}
Due to the cancellation in the last step, $\widetilde{\tau}'_i$ no longer depends on $i$.
Since $\widetilde{f}_i' < \widetilde{f}_i$, this introduces a bias.
We will correct for this afterwards using expected values.

At step $l-2$, the next vertex is known to be adjacent to the target, which
gives the refined factor
\begin{align}
\widetilde\tau_{l-2}(n,d,l)
&= \frac{\widetilde f_{l-2}''}
        {\widetilde f_{l-2}''
        +\frac{d-2}{n-3}
          \left((n-l)\widetilde f_{l-2}+1\right)}.
\end{align}

For the final transition to $v^{(e)}_{l-1}$ (step $l-1$), we know that $v^{(e)}_{l-1}$ is a neighbor.
Therefore, there is one neighbor that was certainly visited, one that has value 1, and of the remaining $d-2$ neighbors, $\frac{n-(l-1)-1}{n-3}=\frac{n-l}{n-3}$ have not been visited and have value $\widetilde f'_{l-1}$:
\begin{align}
  \widetilde\tau_{l-1}(n,d,l)
  &= \frac{1}{1+(d-2)\frac{n-l}{n-3}\widetilde f'_{l-1}}.
\end{align}

We now assemble the approximation for a cycle $c$ of length $l>3$ without relying on an implicit indexing of the cycle vertices.
For every omitted edge $e\in c$, choose one orientation of the resulting open path and write
\[
  \mathcal P_e^\rightarrow
  =(v_0^{(e)},v_1^{(e)},\ldots,v_{l-1}^{(e)}).
\]
Let $\mathcal P_e^\leftarrow$ denote the reverse orientation.
For the forward orientation, define
\begin{align}
\widetilde\rho_e^{\rightarrow}
&:=\widetilde\tau_1(n,d(v_0^{(e)}),\bar d)
  \left[
    \prod_{j=2}^{l-3}
    \widetilde\tau_j(n,d(v_{j-1}^{(e)}))
  \right]\nonumber \cdot{} \widetilde\tau_{l-2}(n,d(v_{l-3}^{(e)}),l)  \widetilde\tau_{l-1}(n,d(v_{l-2}^{(e)}),l),
\end{align}
and define $\widetilde\rho_e^{\leftarrow}$ analogously on the reverse path.
To improve the overall accuracy, we thus get a first approximation
\begin{equation}\label{eq:est-prob}
  \widetilde\rho_c:=\sum_{e\in \mathcal E(c)}\frac{\widetilde\rho_e^{\leftarrow}+\widetilde\rho_e^{\rightarrow}}{2}.
\end{equation}
The average removes the small orientation dependence introduced by the
mean-field substitutions.

For fast path aggregation, we require most factors to be independent of both $i$ and $l$.
We approximate these probabilities further by $\widetilde \tau'_i(n,d)$ to obtain a multiplicative aggregator $\varrho$ and an additive aggregator $\eta$.
We correct the (expected) bias of this approximation with a global correction factor $\gamma$:

\begin{align}
  \widetilde\rho_e^{\rightarrow} &= \widetilde\tau_1(n,d(v_0^{(e)}),\bar d)
  \left[
    \prod_{j=2}^{l-3}
    \widetilde\tau_j(n,d(v_{j-1}^{(e)}))
  \right]\nonumber \cdot{} \widetilde\tau_{l-2}(n,d(v_{l-3}^{(e)}),l)  \widetilde\tau_{l-1}(n,d(v_{l-2}^{(e)}),l) \\
  &\approx \widetilde\tau_1(n,d(v_0^{(e)}),\bar d)\widetilde\tau_{l-2}(n,d(v_{l-3}^{(e)}),l)  \widetilde\tau_{l-1}(n,d(v_{l-2}^{(e)}),l)
        \left[
    \prod_{j=2}^{l-3}
    \widetilde\tau'_i(n,d(v_{j-1}^{(e)}))\right]
    \left[\prod_{j=2}^{l-3}
    \frac{\widetilde\tau_j(n,\bar d)}{\widetilde\tau'_i(n,\bar d)}\right]\\
  &\approx\left[\prod_{j=2}^{l-3}
    \frac{\widetilde\tau_j(n,\bar d)}{\widetilde\tau'_i(n,\bar d)}\right] \frac{\widetilde\tau_1(n,d(v_0^{(e)}),\bar d)  \widetilde\tau_{l-2}(n,d(v_{l-3}^{(e)}),l)  \widetilde\tau_{l-1}(n,d(v_{l-2}^{(e)}),l)}
          {\widetilde\tau'_i(n,d(v_{0}^{(e)}))\widetilde\tau'_i(n,d(v_{l-3}^{(e)}))\widetilde\tau'_i(n,d(v_{l-2}^{(e)}))\widetilde\tau'_i(n,d(v_{l-1}^{(e)}))}
    \prod_{j=0}^{l-1}
    \widetilde\tau'_i(n,d(v_{j}^{(e)}))\\
  &\approx\underbrace{\left[\prod_{j=2}^{l-3}
    \frac{\widetilde\tau_j(n,\bar d)}{\widetilde\tau'_i(n,\bar d)}\right] \frac{\widetilde\tau_{l-2}(n,\bar d,l)}{\widetilde\tau'_i(n,\bar d)} \frac{\widetilde\tau_{l-1}(n,\bar d,l)}{\widetilde\tau'_i(n,\bar d)}}_{\gamma(n,\bar d, l)} \underbrace{\frac{\widetilde\tau_1(n,d(v_0^{(e)}),\bar d)}
          {\widetilde\tau'_i(n,d(v_{0}^{(e)}))\widetilde\tau'_i(n,d(v_{l-1}^{(e)}))}}_{\text{depends on } v_0^{(e)}\text{ and } v_{l-1}^{(e)}}
    \underbrace{\prod_{j=0}^{l-1}
    \widetilde\tau'_i(n,d(v_{j}^{(e)}))}_{\text{only depends on cycle}} =: \rho_e^\mathrm{MF,\rightarrow}
\end{align}
This approximation for the one-way path inclusion probability $\widetilde\rho_e^{\rightarrow}$ consists of the (final) general correction factor $\gamma$, a part depending on the degrees of the first and last vertex in the cycle, and the cycle-dependent (but not path-dependent) product of $\widetilde \tau_i'$ over all vertices.
The approximation for the reversed path $\widetilde\rho_e^{\leftarrow}$ is identical except in the fraction depending on the two vertices, which are reversed.
We can now insert this approximation into \cref{eq:est-prob} to obtain the final approximation $\rho_c^\mathrm{MF}$:
\begin{align}
    \rho_c^\mathrm{MF} &= \sum_{e \in c} \frac{\rho_e^\mathrm{MF,\rightarrow} + \rho_e^\mathrm{MF,\leftarrow}}{2}\\
    &= \sum_{(u,v)\in c} \gamma(n, \bar d, l) \underbrace{\frac{\widetilde\tau_1(n,d(u),\bar d) + \widetilde\tau_1(n,d(v),\bar d)}{2\cdot{}\widetilde\tau'_i(n,d(u))\widetilde\tau'_i(n,d(v))}}_{\eta(n, d(u), d(v))} \prod_{u \in c}
    \underbrace{\widetilde\tau'_i(n,d(u))}_{\varrho(n, d(u))}\\
    &= \gamma(n, \bar d, l) \left(\sum_{(u,v) \in c} \eta(n, d(u), d(v))\right) \prod_{u \in c} \varrho(n, d(u))
\end{align}

The aggregated factors $\gamma, \eta, \varrho$ are functions that only depend on their parameters:
\begin{align}
\gamma(n,\bar d,l)
&:=
  \left[
  \prod_{j=2}^{l-3}
    \frac{\widetilde\tau_j(n,\bar d)}{\widetilde\tau'_i(n,\bar d)}
  \right]
  \frac{\widetilde\tau_{l-2}(n,\bar d,l)}{\widetilde\tau'_i(n,\bar d)}
  \frac{\widetilde\tau_{l-1}(n,\bar d,l)}{\widetilde\tau'_i(n,\bar d)},\\
\eta(n,d_1,d_2)
&:=
  \frac{
    \widetilde\tau_1(n,d_1,\bar d)
    +\widetilde\tau_1(n,d_2,\bar d)}
  {2\cdot \widetilde\tau'_i(n,d_1)\widetilde\tau'_i(n,d_2)},\\
\varrho(n,d_1)&:=\widetilde\tau'_i(n,d_1).
\end{align}

As discussed in \cref{sec:theoretical:complexity}, this approximation can be efficiently calculated using a spanning tree.

\newpage
\section{Example Run on Toy Graph}
\label{app:toy-run}

This section provides an example run to give a better intuition on how the algorithm and its optimizations work.
We will be using the following graph drawn from \erdosrenyi{} ($n=8,p=0.5$):

\includegraphics[width=.4\linewidth]{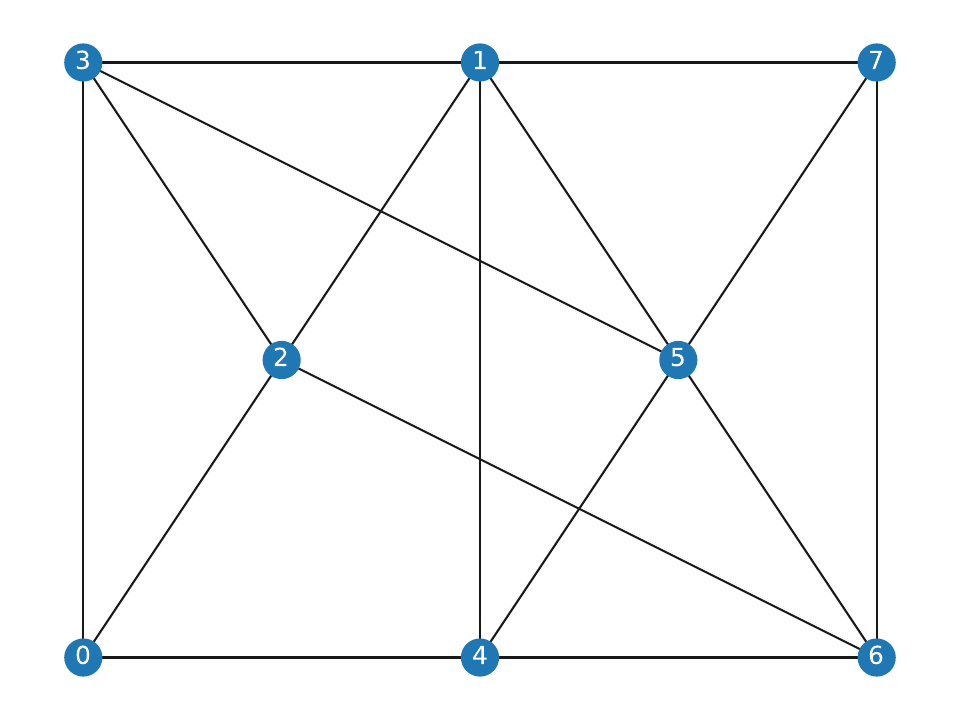}

First, we sample a uniform spanning tree (left: highlighted on graph, right: displayed as rooted spanning tree).

\includegraphics[width=.48\linewidth]{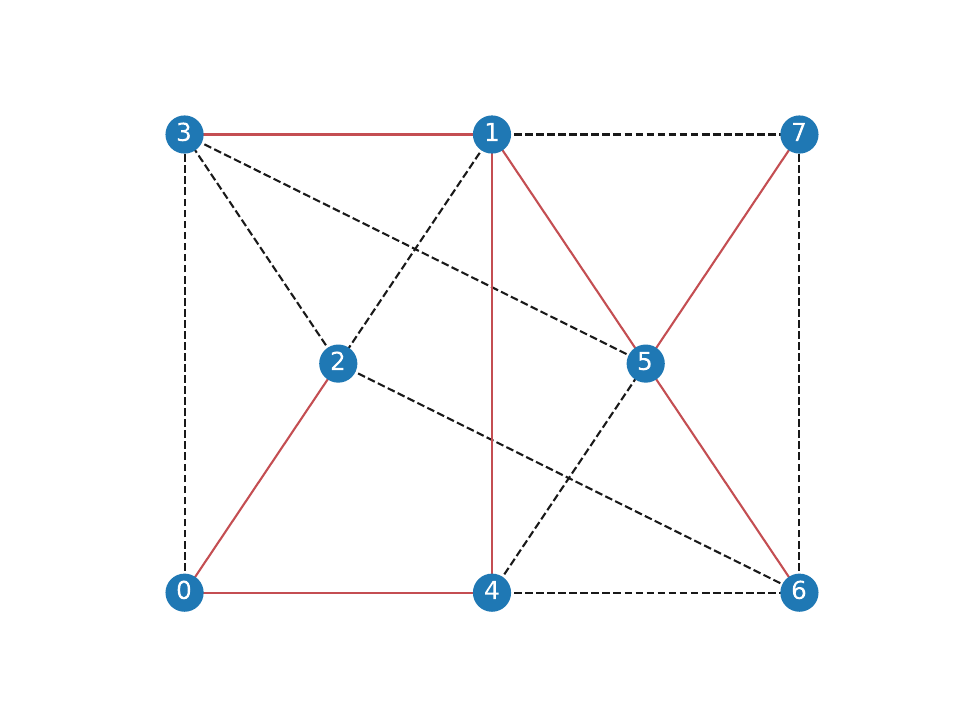}
\hfill
\includegraphics[width=.48\linewidth]{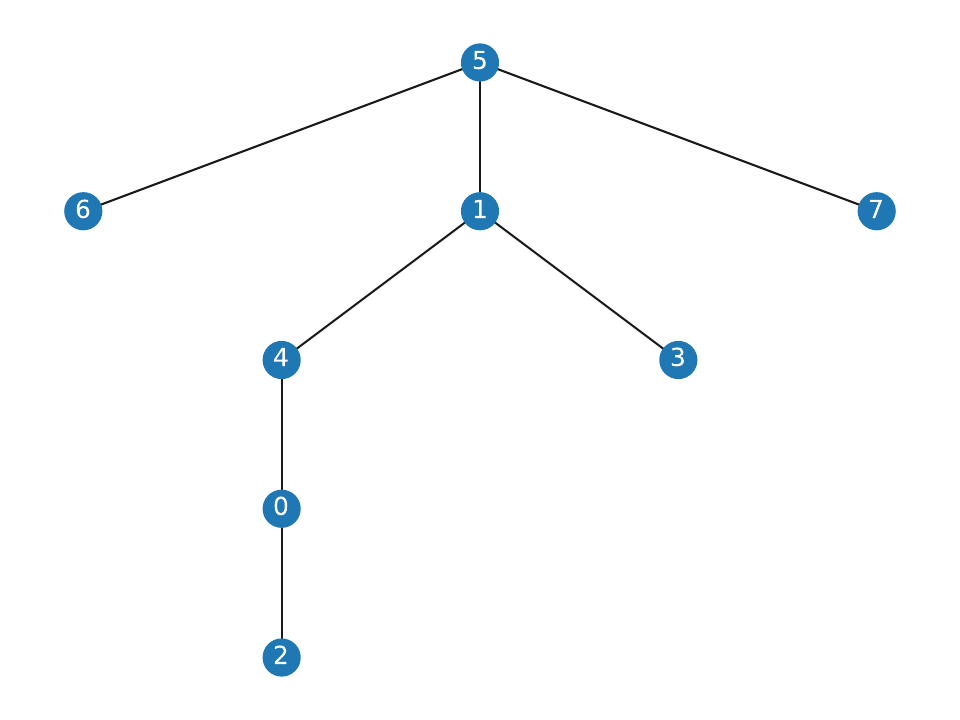}

As explained in \cref{sec:theoretical:complexity}, we can calculate
$\rho_c^{\mathrm{MF}}$ for all cycles $c \in \mathcal C_T$ efficiently.
For this, we calculate $\sigma(r,u), \pi(r,u)$ for all $u \in T$:

\includegraphics[width=.5\linewidth]{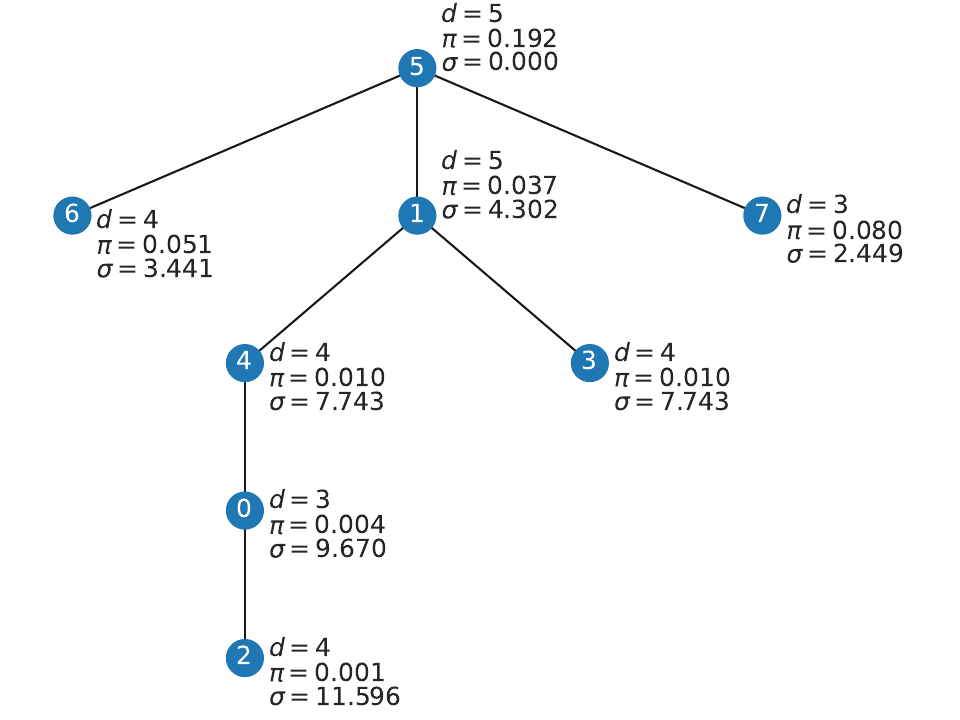} 

Then, we calculate all lowest common ancestors using Tarjan's off-line lca algorithm \cite{tarjan1979applications}.
In our example, we will look at the edge $(0,3)$ with $\lca(0,3)=1$.
The corresponding cycle is $c=(0,3,1,4)$; we can compute its length without iterating over it from the depths of $u,v$, and their lowest common ancestor in the tree.

From this information, we can calculate $\rho_c^{\mathrm{MF}}$ by evaluating
\cref{eq:treecalc_pi,eq:treecalc_sigma}:

\begin{align}
\pi(0,3) &= \frac{\pi(r,0) \pi(r,3)
  \varrho(n,d(\lca(0,3)))}{\pi(r,\lca(0,3))^{2} }
  =\frac{0.0041 \cdot 0.0097 \cdot 0.19}{0.037^2 } = 0.0055\\
    \sigma(0,3) &= \sigma(r,0)+\sigma(r,3)-2\sigma(r,\lca(0,3)) = 9.67+7.74-2\cdot 4.30 = 8.81\\
    \rho_c^{\mathrm{MF}} &= \gamma(n,\bar d,l)
    \cdot{} (\eta(n,d(0),d(3)) + \sigma(0,3)) \cdot{} \pi(0,3)
    = 3.28 \cdot (1.93 + 8.81) \cdot 0.0055 = 0.195
\end{align}

For comparison, the exact cycle inclusion probability is $\rho_c = 0.210$.

\section{Additional Plots}%
\begin{figure}[p]
    \centering
        \includegraphics[width=\linewidth]{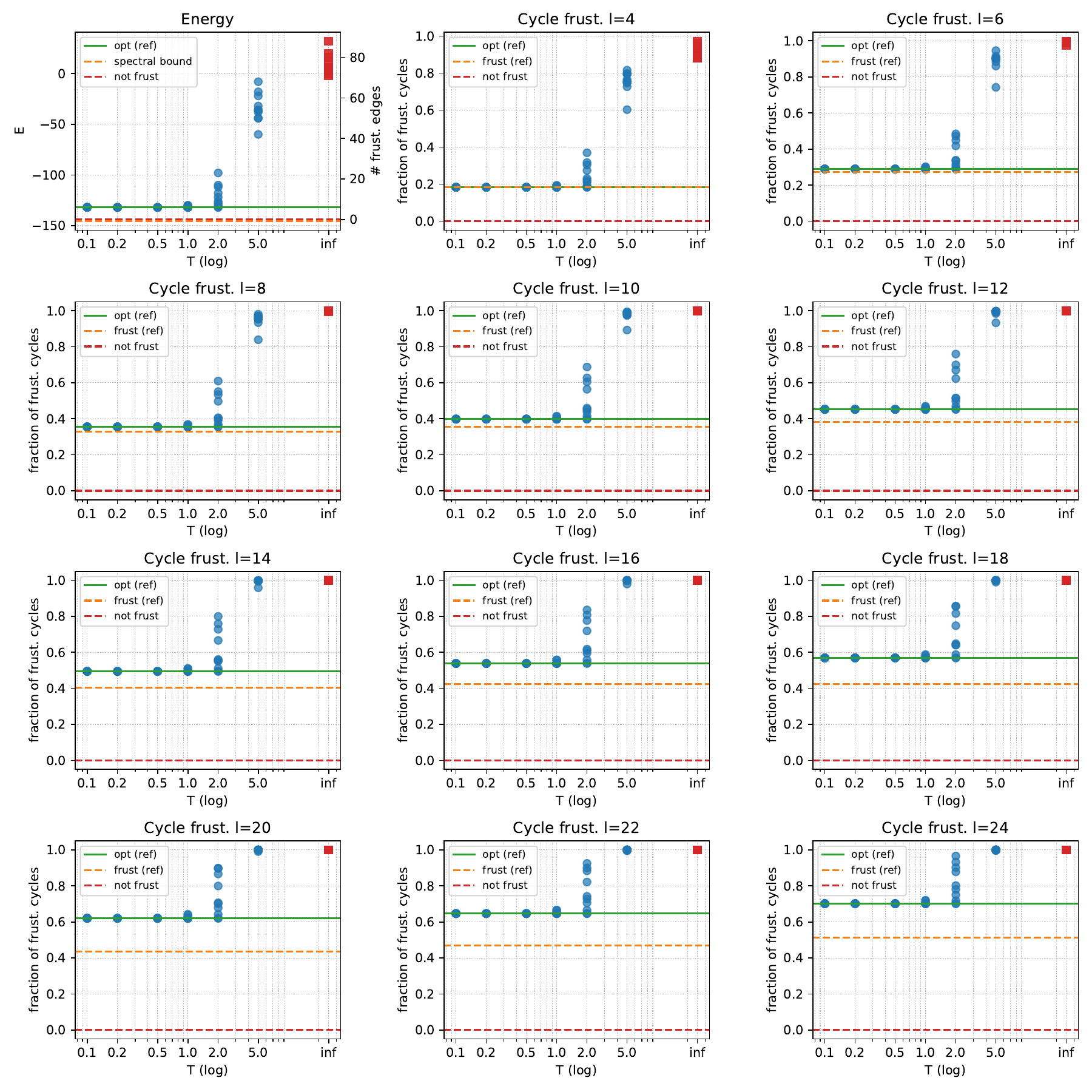}
    \caption{Energy and cycle-level frustration on a $4 \times 4 \times 4$ grid graph with $\mathbb P(J_{i,j}=-1)=0.05$. Cycle-level frustration is estimated with MCHT using the exact $\rho_c$ and $s=1000$ trees. Dots represent temperature-indexed spin configurations and red squares uniformly random configurations. Horizontal lines show values for a non-unique optimum, the relevant energy or cycle-level lower bound, and zero frustration. The cycle-level bound is informative for short lengths and becomes loose for longer cycles; it is not directly comparable in magnitude with the spectral energy bound.}\label{fig:spin-grid}
\end{figure}

\begin{figure}[p]
        \includegraphics[width=\linewidth]{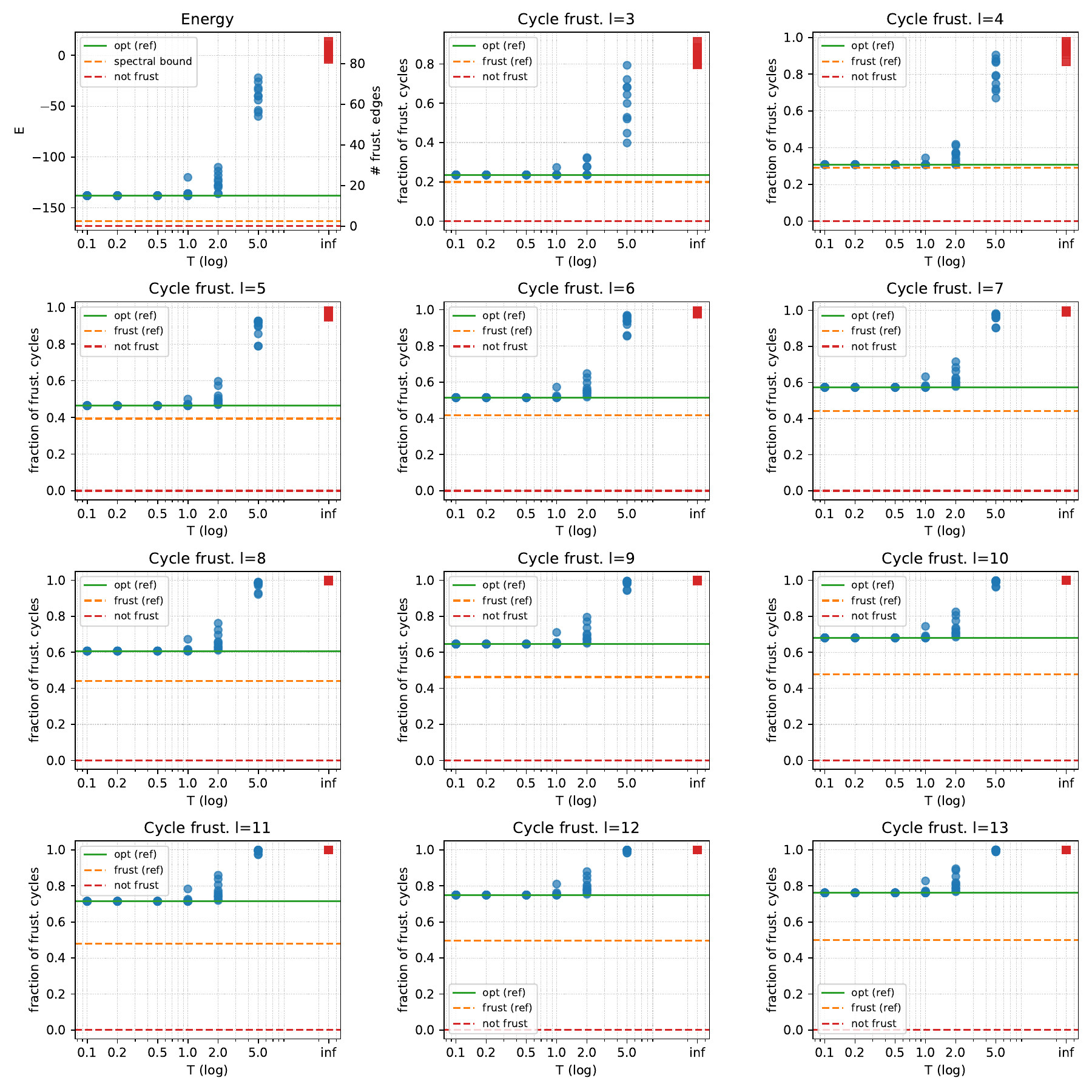}
    \caption{Energy and cycle-level frustration on an \erdosrenyi{} graph with $n=60$, $p=0.1$, and $\mathbb P(J_{i,j}=-1)=0.1$. Plot elements and estimation are as in \cref{fig:spin-grid}. The cycle-level bound is informative for short lengths and becomes loose for longer cycles; it is not directly comparable in magnitude with the spectral energy bound.}\label{fig:spin-er}
\end{figure}

\begin{figure}[p]
        \includegraphics[width=\linewidth]{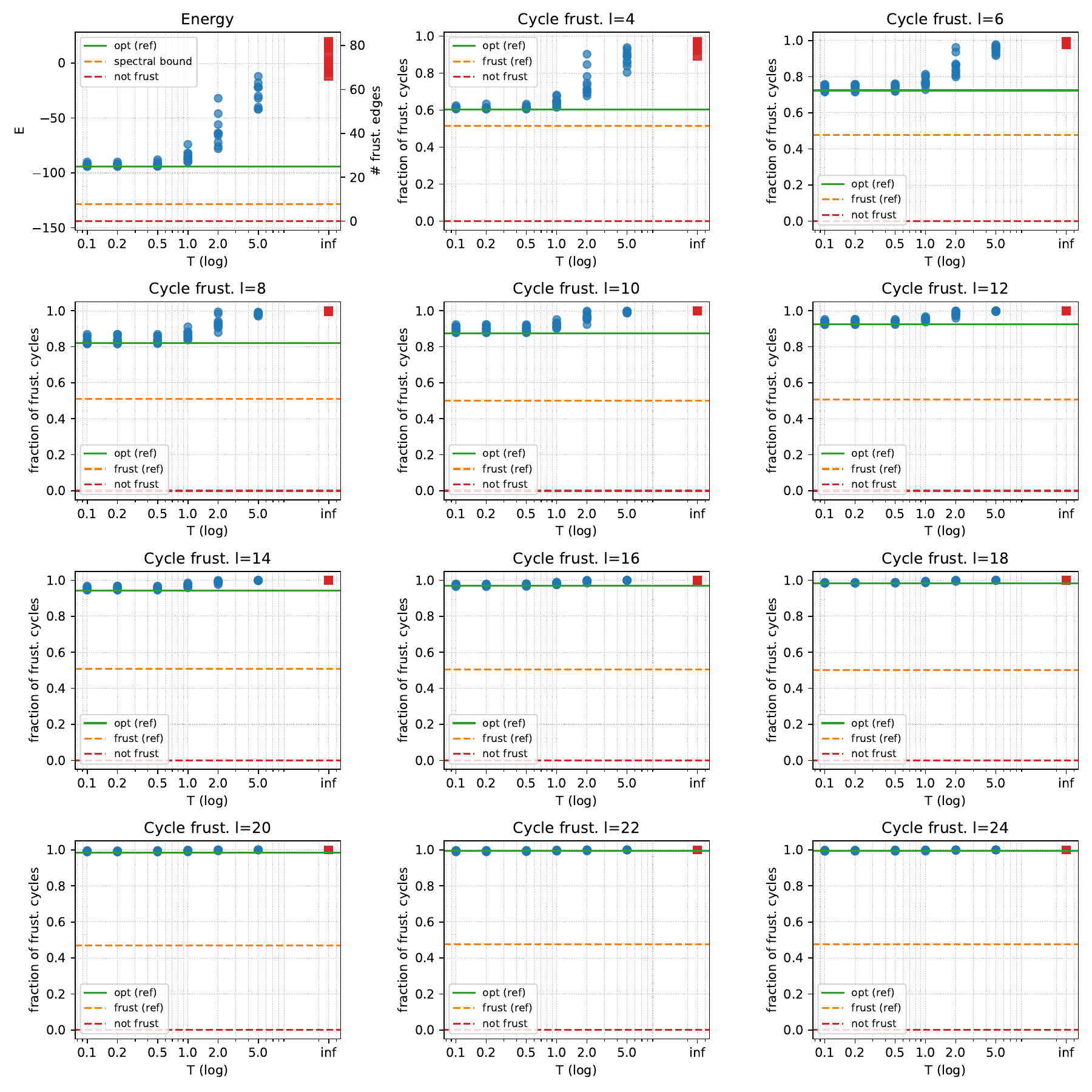}
    \caption{Energy and cycle-level frustration on a $4 \times 4 \times 4$ grid graph with $\mathbb P(J_{i,j}=-1)=0.2$. Plot elements and estimation are as in \cref{fig:spin-grid}. In this regime the cycle-level lower bound is loose.}\label{fig:spin-grid2}
\end{figure}

\begin{figure}[p]
        \includegraphics[width=\linewidth]{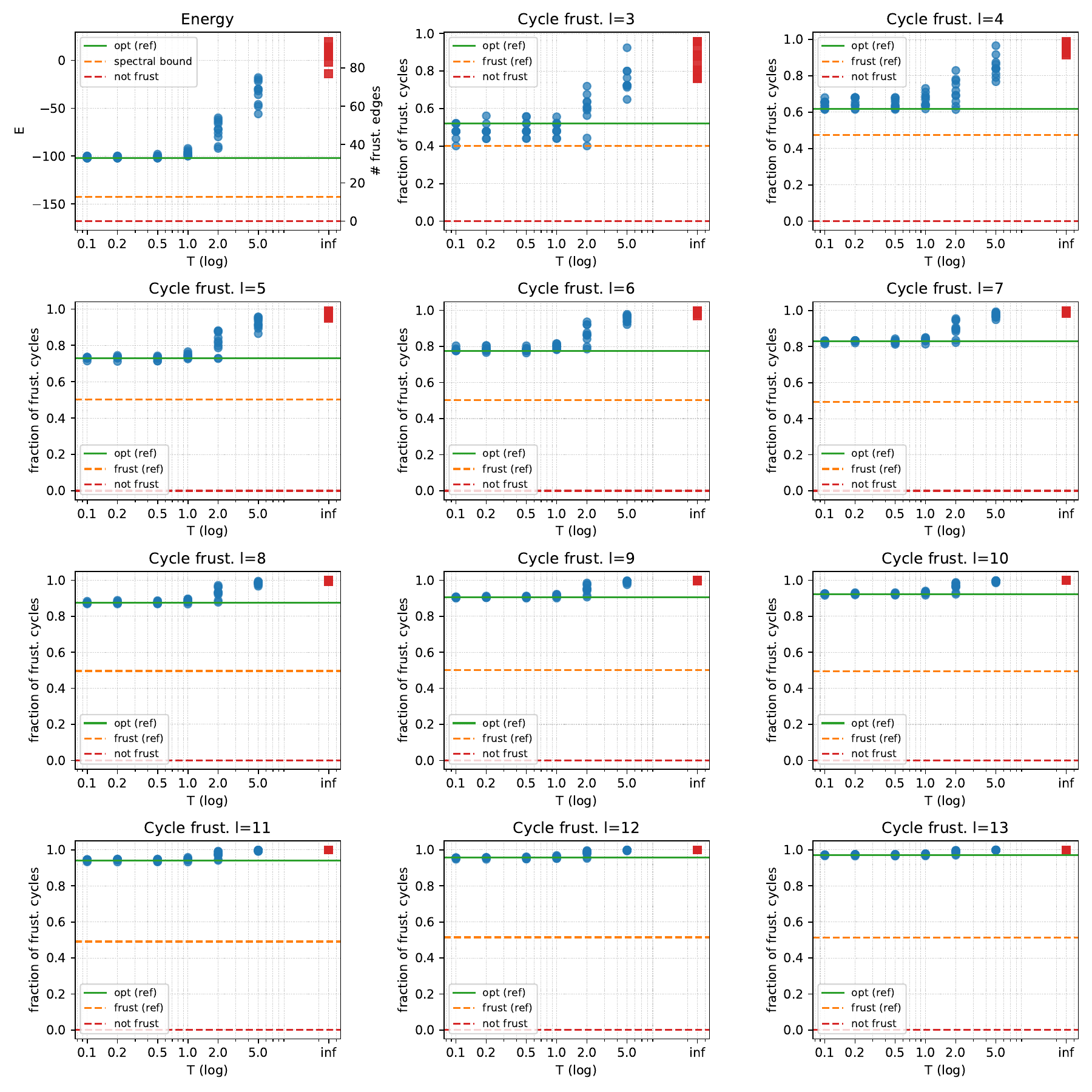}
    \caption{Energy and cycle-level frustration on the same \erdosrenyi{} graph as in \cref{fig:spin-er} ($n=60$, $p=0.1$) with $\mathbb P(J_{i,j}=-1)=0.3$. Plot elements and estimation are as in \cref{fig:spin-grid}. In this regime the cycle-level lower bound is loose.}\label{fig:spin-er2}
\end{figure}

\begin{figure}[p]
    \centering
    \includegraphics[width=\linewidth]{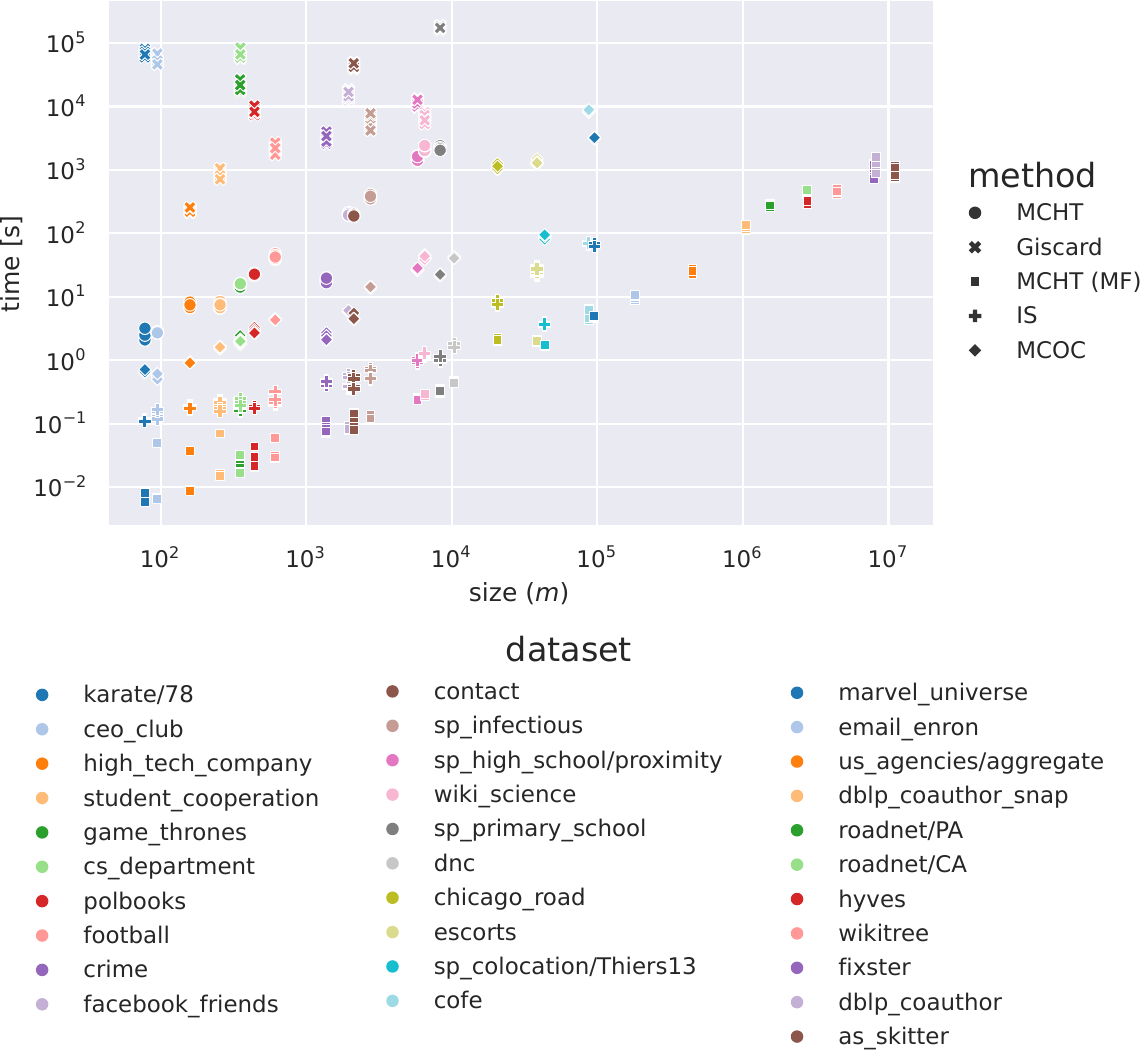}
    \caption{Runtime on selected real-world datasets (retrieved from \textit{Netzschleuder}, \url{https://networks.skewed.de}, same naming convention), ordered by number of edges ($m$); estimation run from $100$ sampled trees (Giscard: $100$ subgraphs of size $l_0=25$); ten runs per graph; exact probability $\rho_c$ unless noted as MF. For disconnected networks, we used the largest connected component. The sample includes some of the largest networks available on Netzschleuder. This shows that our algorithm is performant enough for practical applications on a large number of real-world datasets.}
    \label{fig:time-rw}
\end{figure}

\begin{figure}[p]
    \begin{subfigure}{.49\linewidth}
        \centering
        \includegraphics[width=\linewidth]{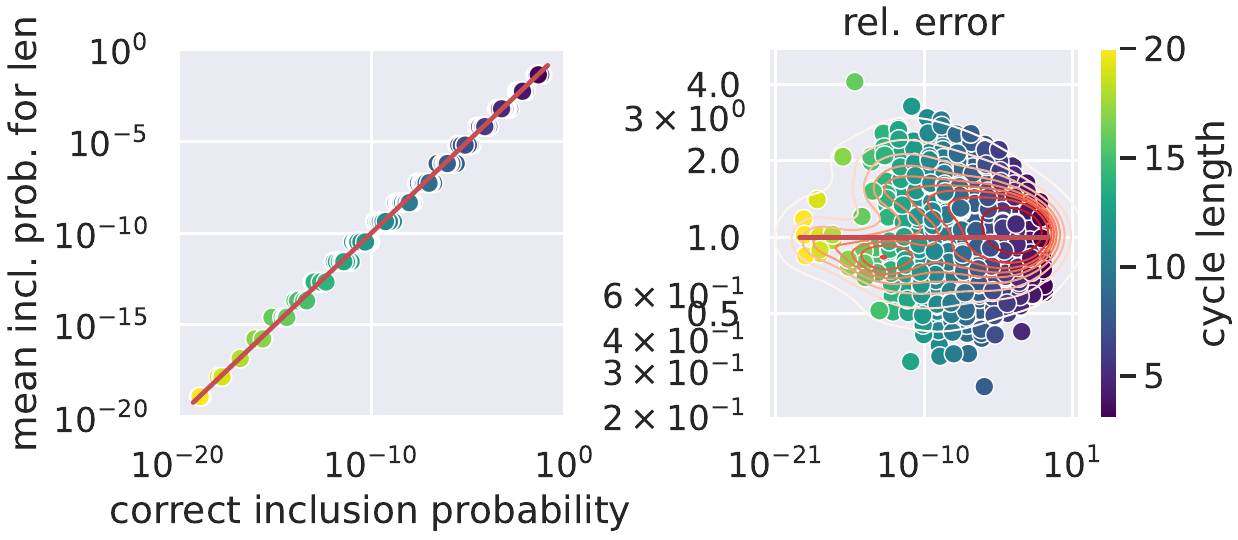}
        \caption{Evaluation on the same cycles and graph sampled from ER with $n=30, p=0.5$ as \cref{fig:estimation_both} (a).}
    \end{subfigure}
    \hfill
    \begin{subfigure}{.49\linewidth}
        \centering
        \includegraphics[width=\linewidth]{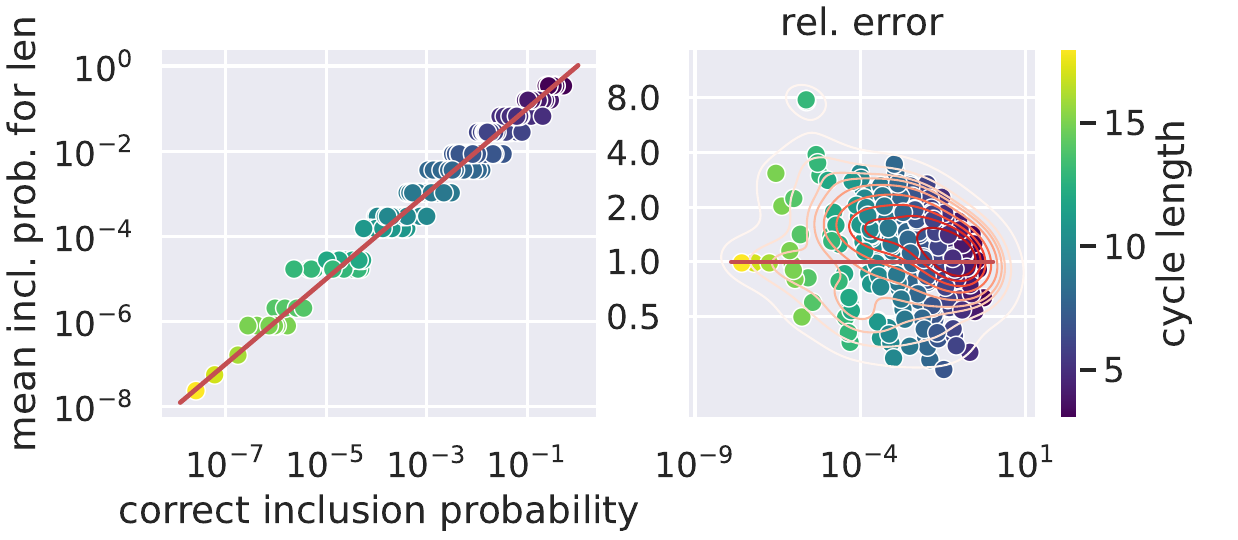}
        \caption{Evaluation on the same cycles and graph sampled from ER with $n=30, p=0.15$ as \cref{fig:estimation_both} (b).}
    \end{subfigure}
        \caption{The approximation error of a hypothetical approximation that assigns each cycle the mean probability of all cycles with the same length, for a moderate configuration (a) and a sparse configuration (b). The estimation error is significantly larger than that of our approach, i.e., our approach detects significantly more than just the cycle length.}
        \label{fig:prob_mean_per_len}
\end{figure}

\begin{figure}[p]
    \begin{subfigure}{.49\linewidth}
        \centering
        \includegraphics[width=\linewidth]{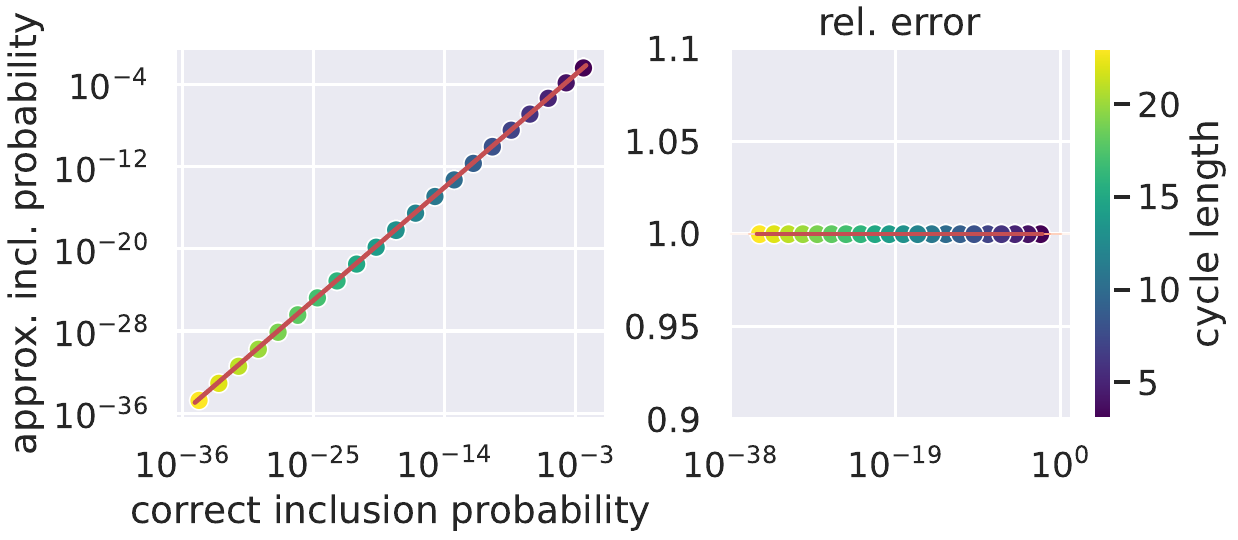}
        \caption{Complete graph $K_{50}$.}
    \end{subfigure}
    \hfill
    \begin{subfigure}{.49\linewidth}
        \centering
        \includegraphics[width=\linewidth]{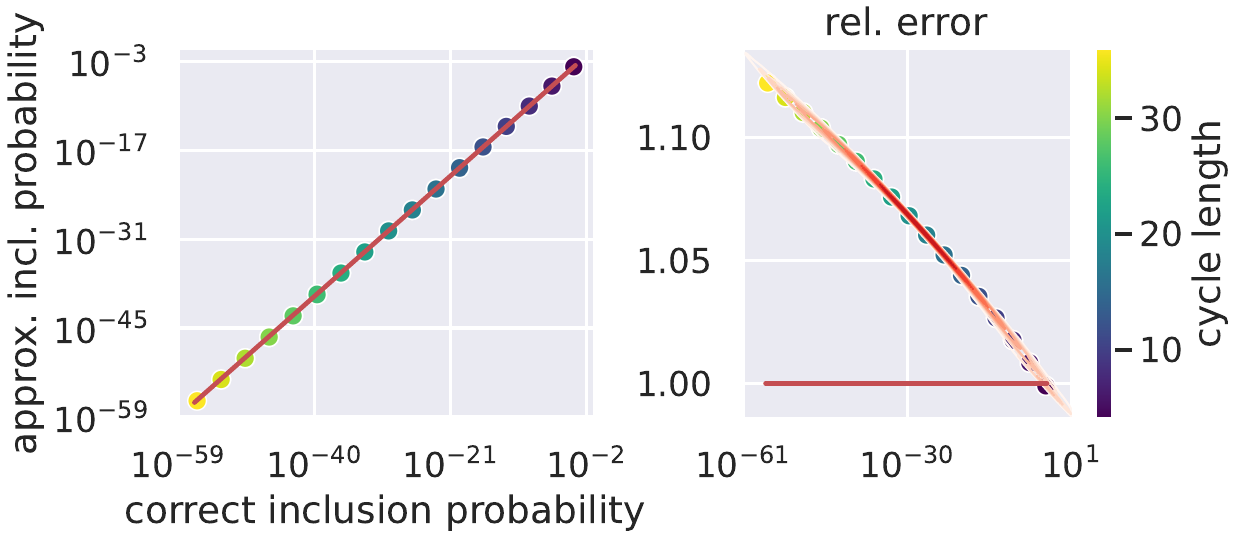}
        \caption{Complete bipartite graph $K_{50,50}$.}
    \end{subfigure}
    \begin{subfigure}{.49\linewidth}
        \centering
        \includegraphics[width=\linewidth]{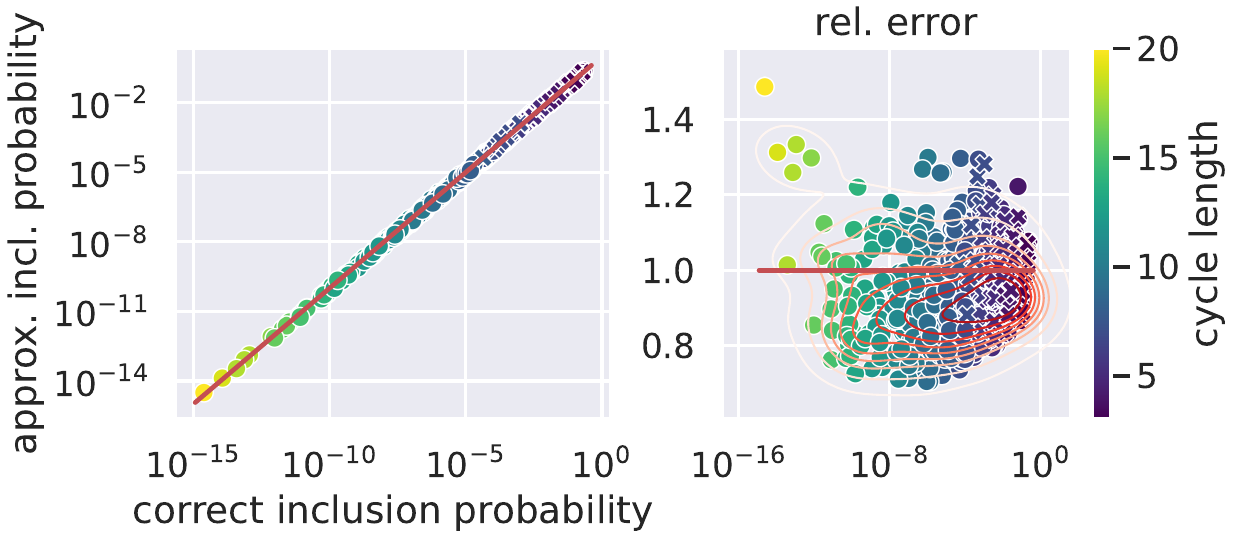}
        \caption{Homophilic SBM \cite{holland1983stochastic} (two blocks, $n=30, p=0.4, q=0.2$).}
        \label{fig:est:sbm-homophilic}
    \end{subfigure}
    \hfill
    \begin{subfigure}{.49\linewidth}
        \centering
        \includegraphics[width=\linewidth]{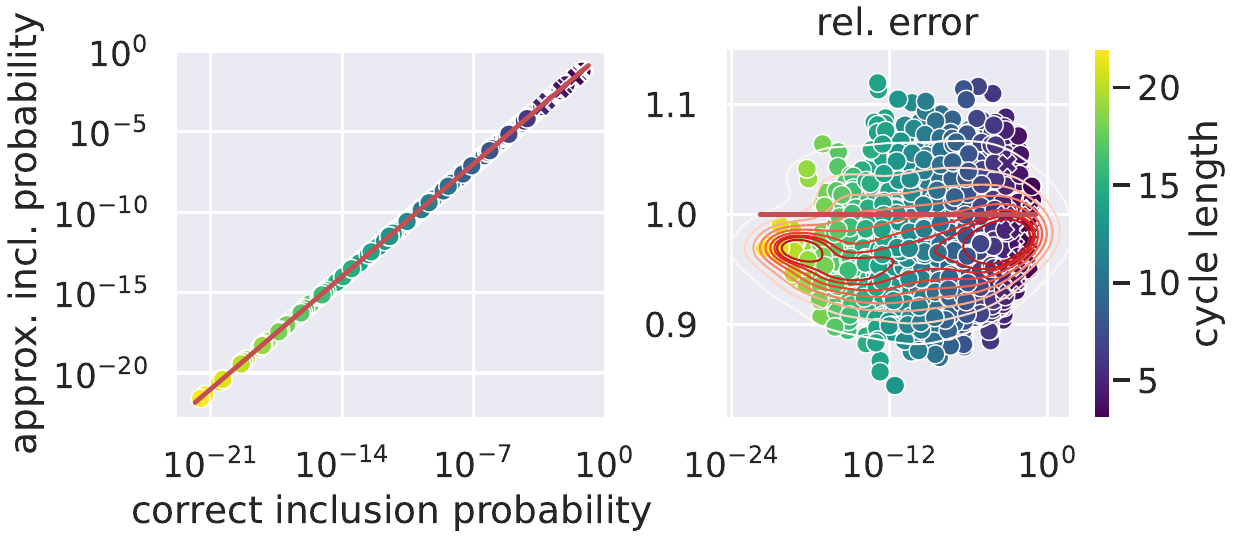}
        \caption{Heterophilic SBM (two blocks, $n=50, p=0.2, q=0.4$).}
        \label{fig:est:sbm-heterophilic}
    \end{subfigure}
    \begin{subfigure}{.49\linewidth}
        \centering
        \includegraphics[width=\linewidth]{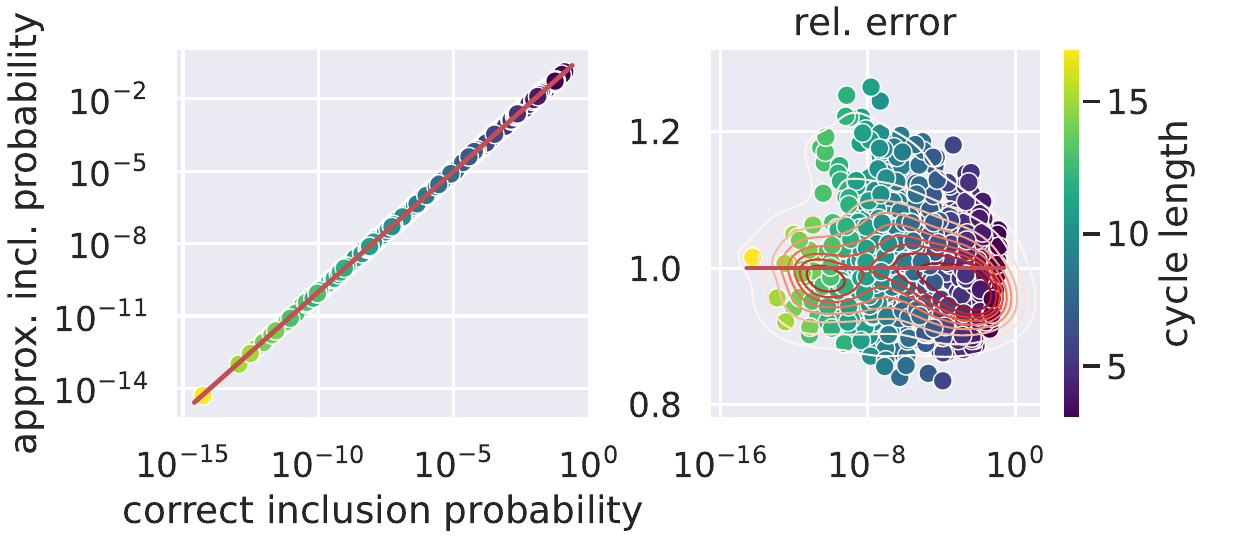}
        \caption{Barab\'asi--Albert \cite{barabasi1999emergence} graph with $30$ vertices, each new vertex attaching with $7$ edges.}
    \end{subfigure}
    \hfill
    \begin{subfigure}{.49\linewidth}
        \centering
        \includegraphics[width=\linewidth]{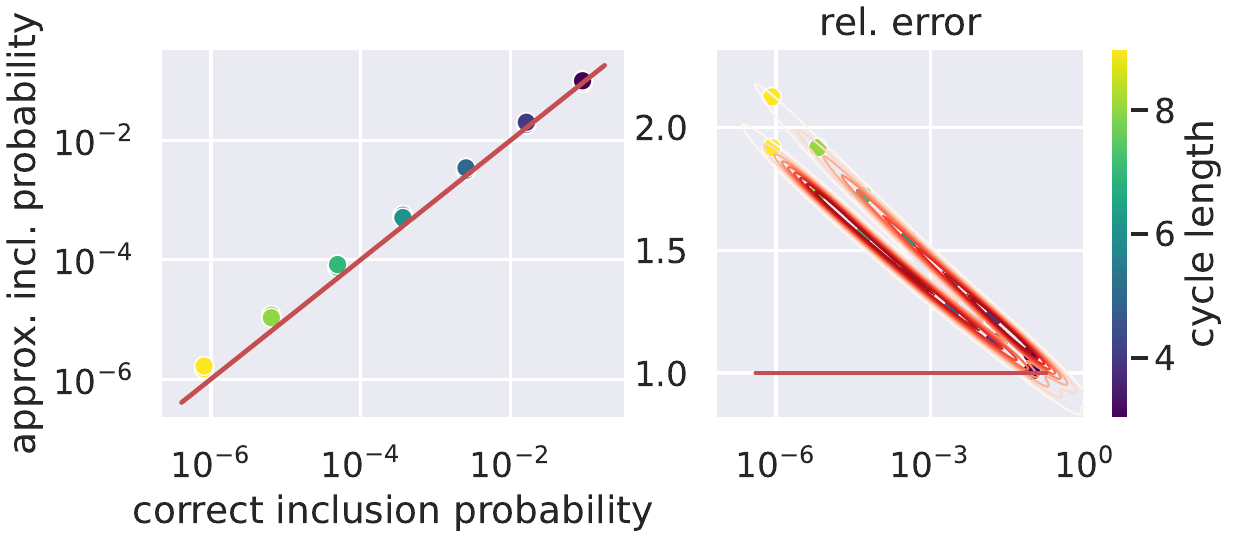}
        \caption{Barbell graph consisting of two $10$-vertex cliques connected
        by a path through three additional vertices.}
    \end{subfigure}
    \begin{subfigure}{.49\linewidth}
        \centering
        \includegraphics[width=\linewidth]{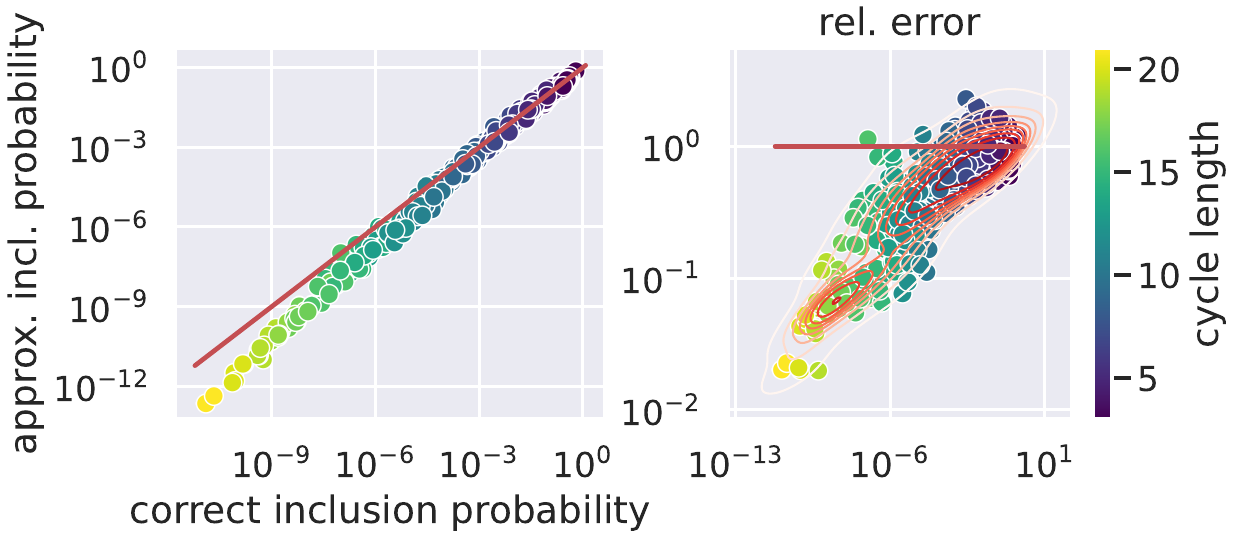}
        \caption{Delaunay triangulation of 50 uniformly random points in two-dimensional Euclidean space.}
    \end{subfigure}
    \hfill
    \begin{subfigure}{.49\linewidth}
        \centering
        \includegraphics[width=\linewidth]{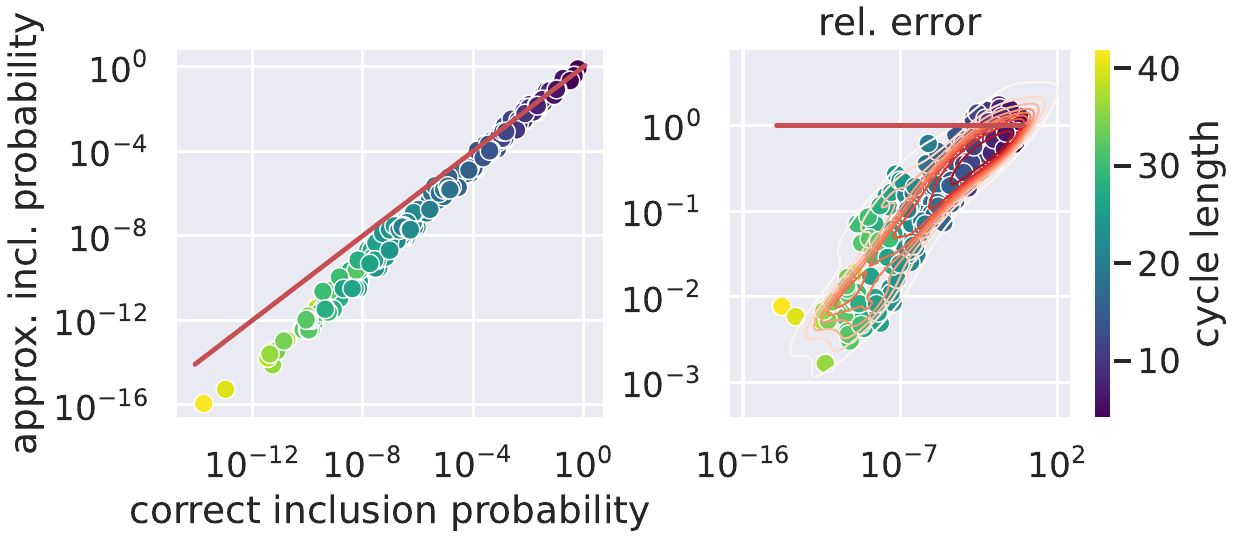}
        \caption{10 by 10 grid graph.}
    \end{subfigure}
        \caption{\textbf{Approximation error on synthetic graphs}, for cycles exposed by ten USTs (as in \cref{fig:estimation_both}). The
        approximation is exact on the complete graph and remains accurate on
        several well-mixed models.  Its systematic error grows when local
        connectivity is a poor description of global geometry.  The barbell
        graph introduces a bottleneck, while the Delaunay triangulation and
        grid have strong low-dimensional structure and large diameter.  In
        these cases, vertices with the same degree can have different harmonic
        relationships to the target, which the mean-field approximation cannot
        represent.}
        \label{fig:appr_synth}
\end{figure}

\begin{figure}[p]
    \begin{subfigure}{.49\linewidth}
        \centering
        \includegraphics[width=.92\linewidth]{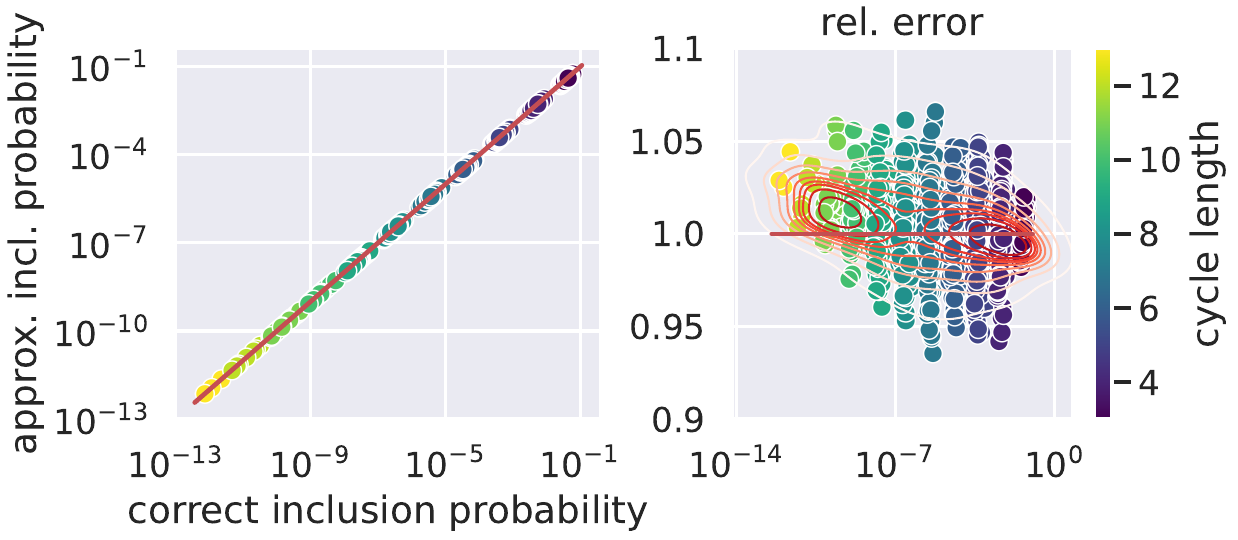}
        \caption{High Tech Company.}
    \end{subfigure}
    \hfill
    \begin{subfigure}{.49\linewidth}
        \centering
        \includegraphics[width=.92\linewidth]{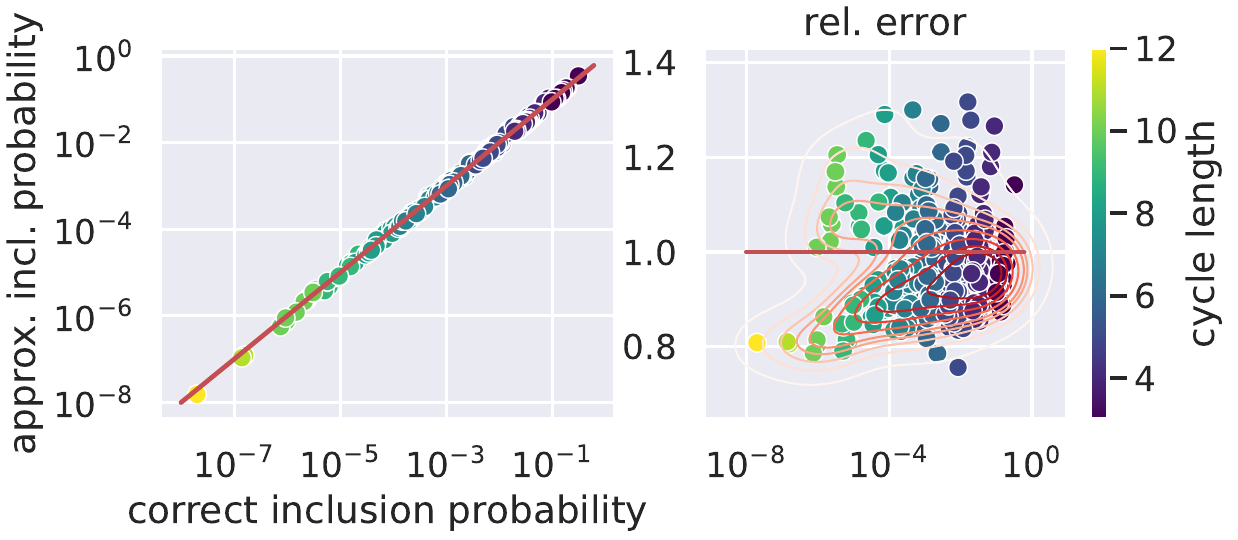}
        \caption{New Guinea Tribes.}
    \end{subfigure}
    \begin{subfigure}{.49\linewidth}
        \centering
        \includegraphics[width=.92\linewidth]{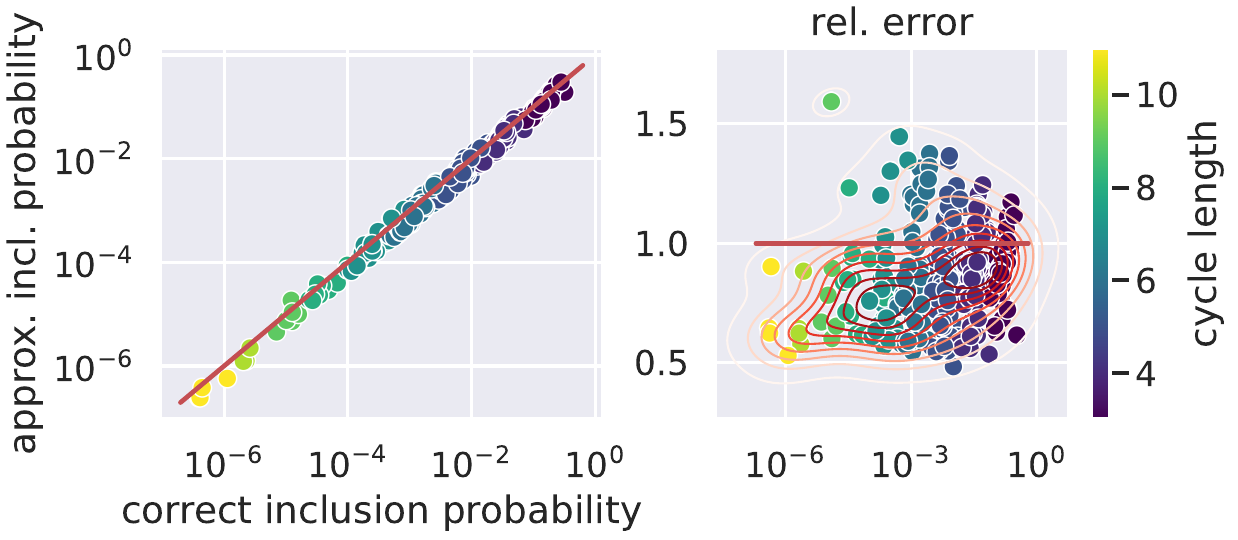}
        \caption{November 17 Terror Cell Network.}
    \end{subfigure}
    \hfill
    \begin{subfigure}{.49\linewidth}
        \centering
        \includegraphics[width=.92\linewidth]{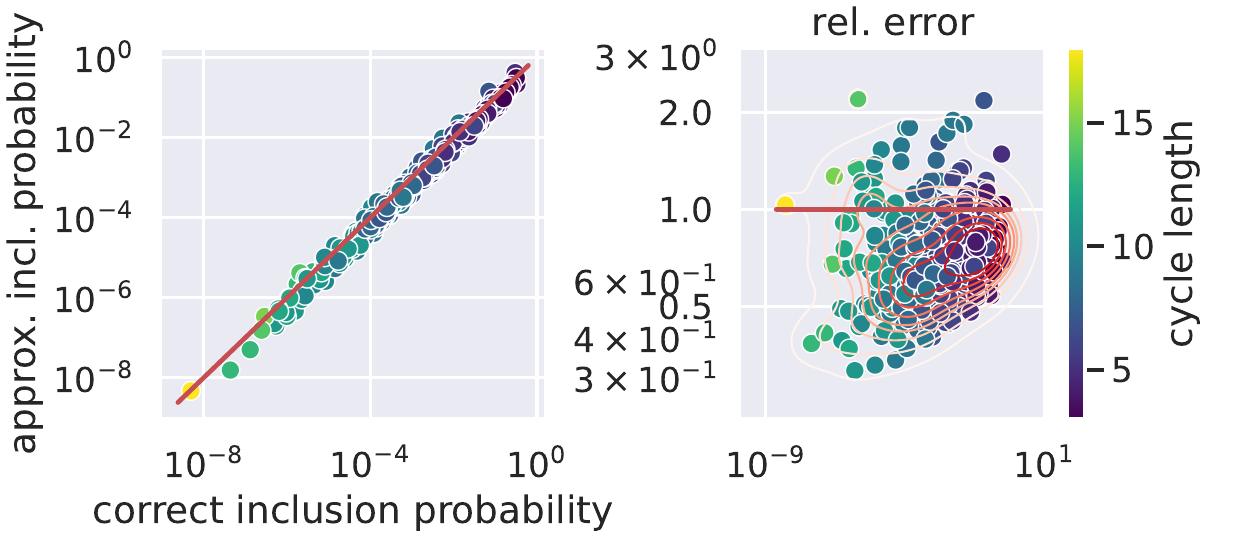}
        \caption{Dutch Criticism.}
    \end{subfigure}
    \begin{subfigure}{.49\linewidth}
        \centering
        \includegraphics[width=.92\linewidth]{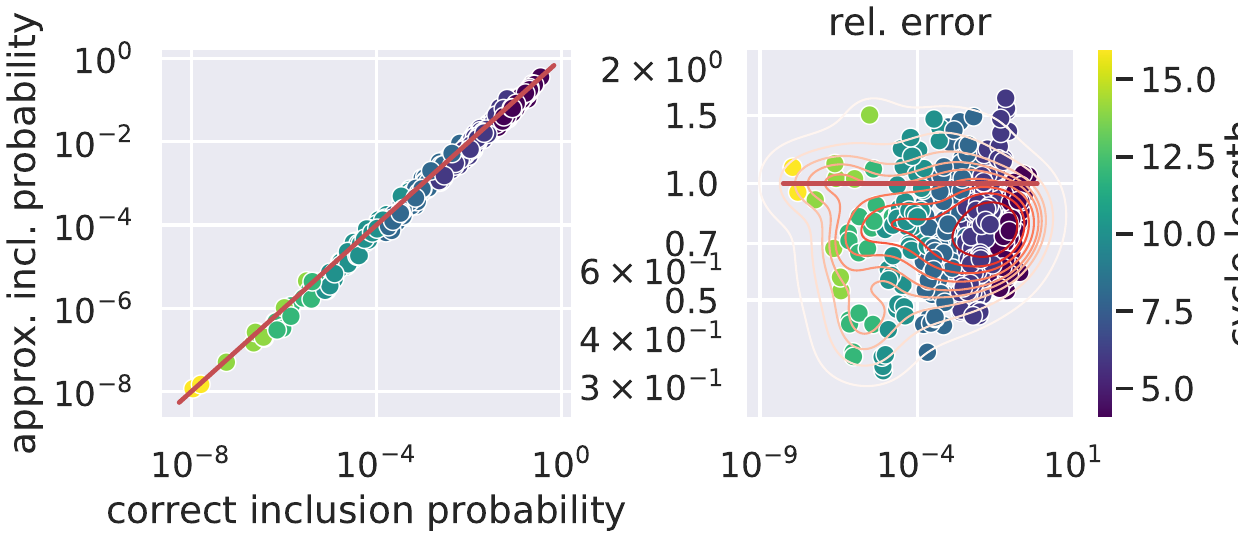}
        \caption{CEO Club.}
    \end{subfigure}
    \hfill
    \begin{subfigure}{.49\linewidth}
        \centering
        \includegraphics[width=.92\linewidth]{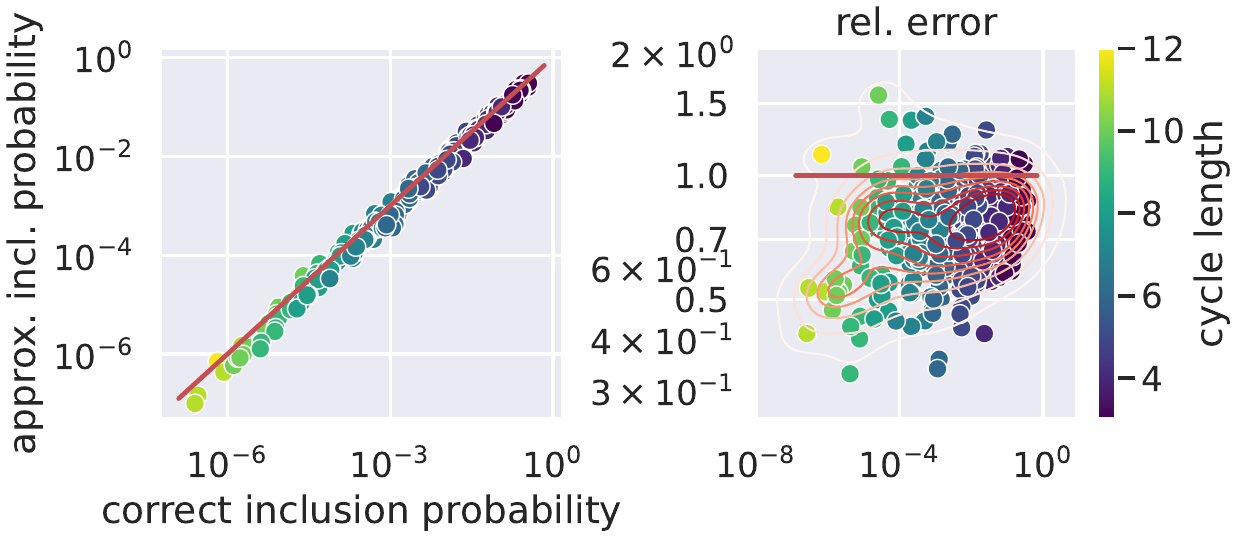}
        \caption{Montreal.}
    \end{subfigure}
    \begin{subfigure}{.49\linewidth}
        \centering
        \includegraphics[width=.92\linewidth]{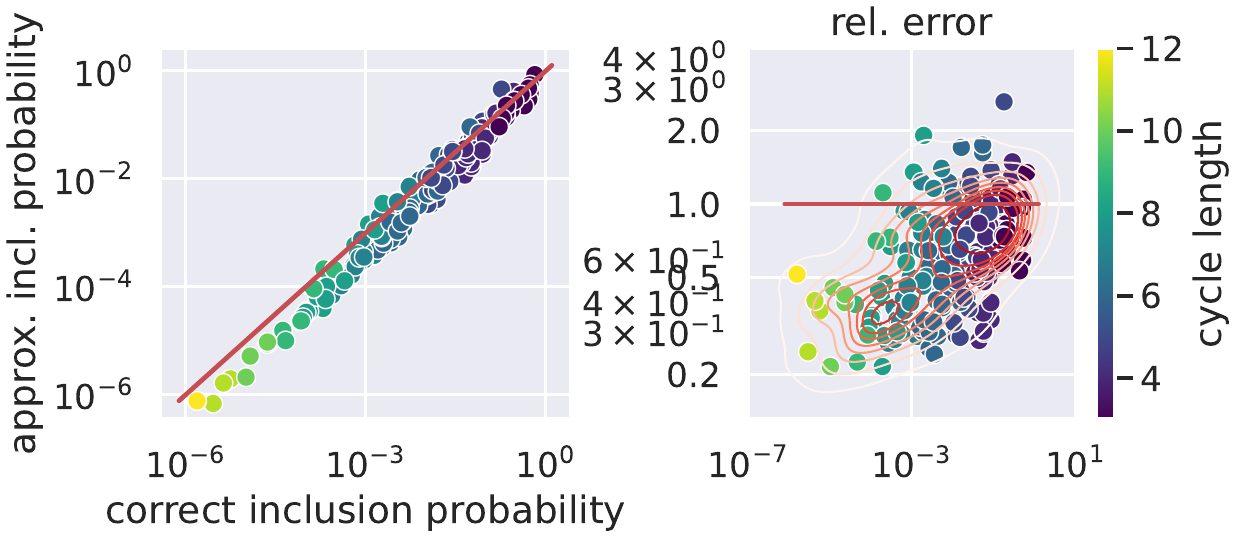}
        \caption{Zachary's Karate Club \cite{zachary1977information}.}
        \label{fig:est:karate}
    \end{subfigure}
    \hfill
    \begin{subfigure}{.49\linewidth}
        \centering
        \includegraphics[width=.92\linewidth]{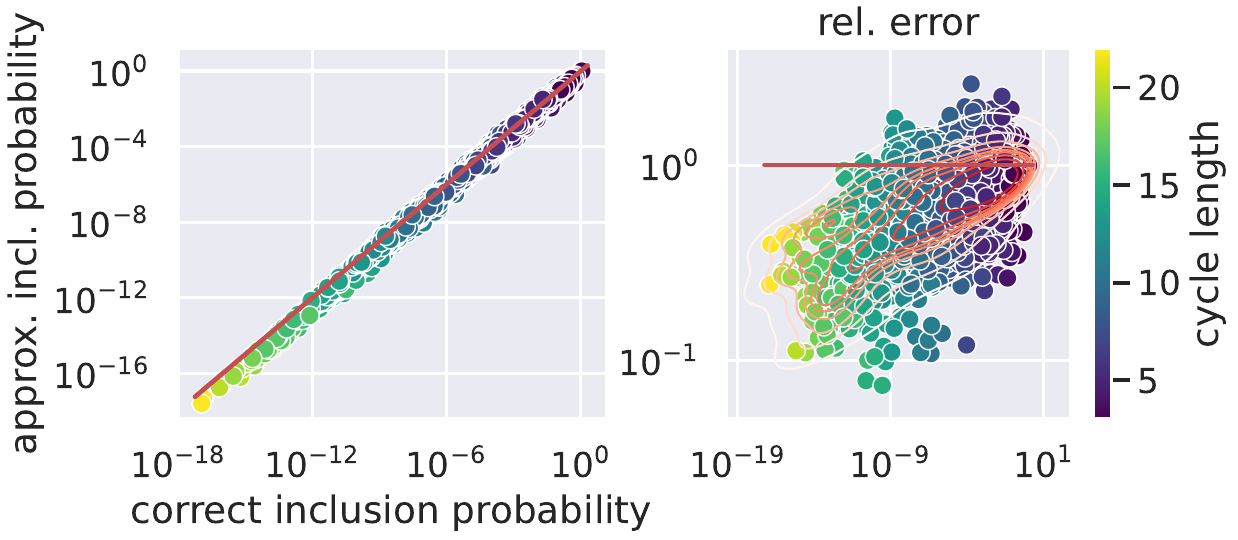}
        \caption{Game [of] Thrones.}
    \end{subfigure}
    \begin{subfigure}{.49\linewidth}
        \centering
        \includegraphics[width=.92\linewidth]{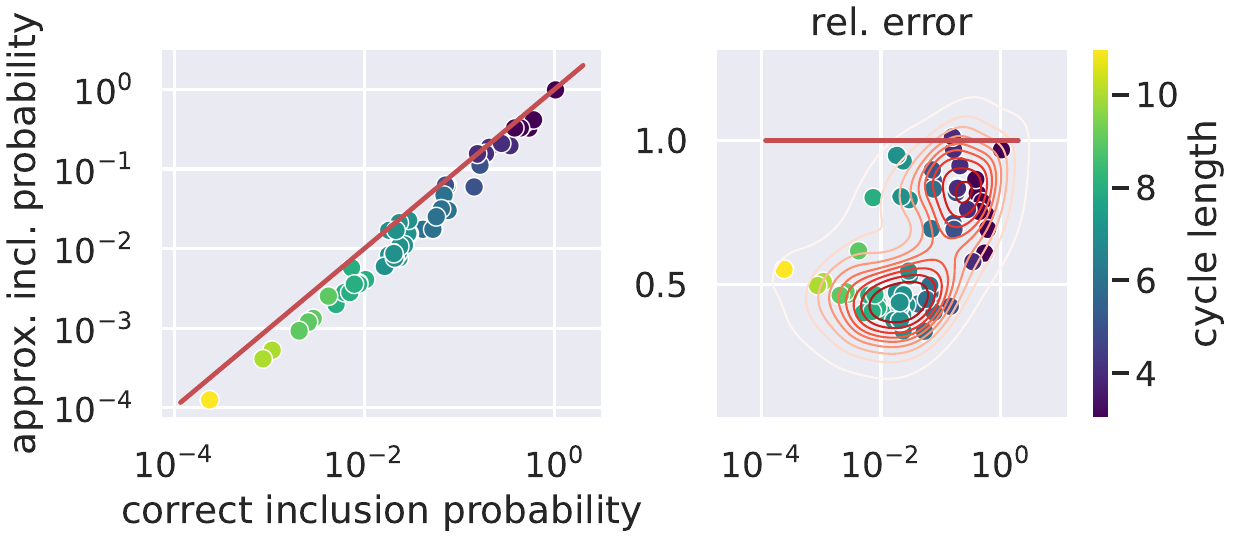}
        \caption{Florentine Families.}
    \end{subfigure}
    \hfill
    \begin{subfigure}{.49\linewidth}
        \centering
        \includegraphics[width=.92\linewidth]{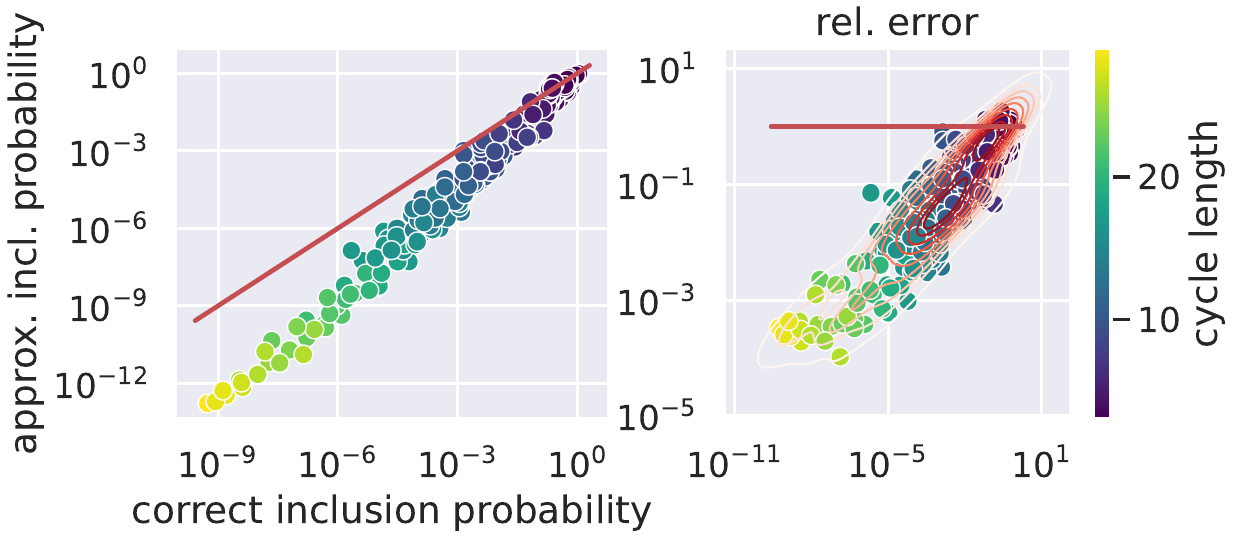}
        \caption{Student Cooperation.}
    \end{subfigure}
    \hfill
        \caption{Accuracy of the approximate cycle inclusion probability
        $\rho_c^{\mathrm{MF}}$ for cycles exposed by ten USTs on real-world networks from \textit{Netzschleuder}
        (\url{https://networks.skewed.de}), unless noted otherwise; largest connected component.  The
        approximation generally has small bias and moderate variance.  The more
        systematic errors in (g)--(i) coincide with comparatively large graph
        diameter; (j), a nearly planar network, is a notable exception.}
        \label{fig:appr_realworld}
\end{figure}

Evaluation code to generate these plots is available at \repolink.
The full spin-system results corresponding to
\cref{fig:spin-glass-good} are shown in
\cref{fig:spin-grid,fig:spin-er,fig:spin-grid2,fig:spin-er2}.
\Cref{fig:time-rw} extends the runtime comparison to real-world networks.
The probability-approximation diagnostics are collected in
\cref{fig:prob_mean_per_len,fig:appr_synth,fig:appr_realworld}.

